\documentclass{elsart}
\usepackage{natbib}
\usepackage{graphicx}

\begin{document}

\newcommand{\s}{\it{s}}
\newcommand{\hl}{$^{4}$He}
\newcommand{\cd}{$^{12}$C}
\newcommand{\ct}{$^{13}$C}
\newcommand{\nq}{$^{14}$N}
\newcommand{\nv}{$^{22}$Ne}
\newcommand{\pd}{$^{208}$Pb}
\newcommand{\bc}{$^{136}$Ba}
\newcommand{\tc}{$^{124}$Te}
\newcommand{\tcc}{$^{99}$Tc}
\newcommand{\rt}{$^{136}$Ba/$^{124}$Te}
\newcommand{\ctan}{$^{13}$C($\alpha$,n)$^{16}$O}
\newcommand{\nean}{$^{22}$Ne($\alpha$,n)$^{25}$Mg}
\newcommand{\neag}{$^{22}$Ne($\alpha$,$\gamma$)$^{26}$Mg}
\newcommand{\oag}{$^{18}$O($\alpha$,$\gamma$)$^{22}$Ne}
\newcommand{\cdpg}{$^{12}$C(p,$\gamma$)$^{13}$N($\beta^+\nu$)$^{13}$C}
\newcommand{\ctp}{$^{13}$C(p,$\gamma$)$^{14}$N}
\newcommand{\cdag}{$^{12}$C($\alpha$,$\gamma$)$^{16}$O}
\newcommand{\cqp}{$^{14}$C(p,$\gamma$)$^{15}$N}
\newcommand{\mvp}{$^{24}$Mg(p,$\gamma$)$^{25}$Al}
\newcommand{\nqnp}{$^{14}$N(n,p)$^{14}$C}
\newcommand{\nqpg}{$^{14}$N(p,$\gamma$)$^{15}$O}
\newcommand{\odna}{$^{17}$O(n,$\alpha$)$^{14}$C}
\newcommand{\alnp}{$^{26}$Al(n,p)$^{26}$Mg}
\newcommand{\alna}{$^{26}$Al(n,$\alpha$)$^{23}$Na}
\newcommand{\stna}{$^{33}$S(n,$\alpha$)$^{30}$Si}
\newcommand{\ctnp}{$^{35}$Cl(n,p)$^{35}$S}
\newcommand{\csna}{$^{36}$Cl(n,$\alpha$)$^{33}$P}
\newcommand{\asnp}{$^{37}$Ar(n,p)$^{37}$Cl}
\newcommand{\asna}{$^{37}$Ar(n,$\alpha$)$^{34}$S}
\newcommand{\csnp}{$^{36}$Cl(n,p)$^{36}$S}
\newcommand{\anna}{$^{39}$Ar(n,$\alpha$)$^{36}$S}
\newcommand{\canp}{$^{41}$Ca(n,p)$^{41}$K}
\newcommand{\cana}{$^{41}$Ca(n,$\alpha$)$^{38}$Ar}
\newcommand{\aaa}{3$\alpha$---$>^{12}$C}
\newcommand{\zsb}{$Z_{\odot}$~}
\newcommand{\ms}{$M_{\odot}$}

\runauthor{Straniero, Gallino \& Cristallo}
\begin{frontmatter}

\title{s Process in low-mass Asymptotic Giant Branch Stars}

\author[Teramo,Napoli]{Oscar Straniero}
\author[Torino,Monash]{Roberto Gallino}
\author[Teramo,Napoli]{Sergio Cristallo}

\address[Teramo]{INAF-Osservatorio Astronomico di Collurania and Universita' di Teramo, via M. Maggini, 64100 Teramo, Italy}
\address[Napoli]{INFN-Sezione di Napoli, Complesso Universitario di Monte Sant'Angelo, Via Cintia, 80126 Napoli, Italy}
\address[Torino]{Dipartimento di Fisica Generale, 
Universita' di Torino, INFN-Sezione di Torino, via P. Giuria 1, 10125 Torino, Italy}
\address[Monash]{Centre for Stellar and Planetary Astrophysics, School of Mathematical Sciences, Monash University,
 3800 Victoria, Australia}

\begin{abstract}
The main component of the s process is produced by low mass stars 
(1.5 $\le$  $M/M_\odot$ $\le$ 3),
 when they climb for the
second time the red giant branch and experience a series of He shell flashes called
thermal pulses. During 
the relatively long period ($10^5$ yr) that elapses between two
subsequent thermal pulses, a slow neutron flux is provided by the \ctan~ reaction
taking place within a thin $^{13}$C pocket located in the He-rich and 
C-rich mantel of these stars. 
A second, marginal, neutron burst occurs during the thermal pulse and it is powered by the 
\nean~ reaction. 
We review the present status of the nucleosynthesis models
of low mass AGB stars. The advance in the knowledge of the complex 
coupling between convective mixing and nuclear process, 
which allows the surface enrichment of C and s-process elements,
is presented, together with the hypotheses 
concerning the physical mechanism driving the formation of the 
$^{13}$C pocket. 
In order to illustrate the capabilities and the limits of the theory,
an updated computation of a 2 $M_\odot$ stellar structure
with solar chemical composition is reported. 
This model has been obtained by including
a full nuclear network 
(from H up to Bi, at the termination point of the s-process path) into the stellar evolution code. 
The predicted modification of the surface 
composition occurring during the AGB
evolution is shown.
The new challenge of AGB modeling, namely the study of C-rich and s-rich 
very metal-poor stars, is discussed.    
\end{abstract}
\end{frontmatter}

\section{Introduction}
It was in 1868 when  the Jesuit astronomer Father Angelo Secchi 
first recognized in the peculiar spectrum of 
some red giants the signature of carbon enhancement 
\cite{se1868}.
This new class includes stars belonging to the Disk of the 
Milky Way, called by Secchi "red carbon stars" and nowaday 
named C(N-Type) stars,
that are evolved low mass red giants 
with photospheric C/O $>$ 1.
In the Hertzsprung-Russell diagram, they are located 
near the tip of the asymptotic giant branch (AGB) and
represent the end point
of the evolutionary sequence of a star that starts the AGB as
an M giant (typically C/O $\approx$ 0.5) and 
progressively modifies its surface composition,
passing through the MS, S and C(N) stages.
The internal structure of these giant stars is made of 
three regions:
a compact C-O core, a thin He-rich and C-rich mantel (He intershell)
and an expanded H-rich envelope. 
As usual for red giant stars, the envelope is largely unstable
against convection. 
Carbon is synthesized by the $3\alpha$ reactions that burn at the base of an 
He-rich layer surrounding the core.
The first question for the
theoreticians was the search for a process capable to move the C from the 
deep interior of the star to the surface (\cite{ib83}, \cite{lat89}, \cite{stra97}).
Modern studies 
have clarified that the dredge up of C is due to the combination  
of two distinct convective episodes. The
first is responsible for an efficient mixing of the whole He-rich layer and the second
partially overlaps the zone previously mixed by the first convective episode 
and extends to the stellar surface.   
As a consequence of these two convective episodes, other products of the 
nucleosynthesis occurring within the He-rich mantel,
 besides $^{12}$C, should appear at the surface.
As a matter of fact, MS, S, C(N) and 
post-AGB stars are enriched in s-process elements, 
like  Sr, Y, Zr, and Ba, La, Ce, Nd corresponding to 
the light and heavy s-abundance peaks, 
as early encoded by \cite{b2fh}.  
Detection of unstable $^{99}$Tc, whose half life
is "only" 2.1 $\times$ $10^5$ yr, demonstrates that 
this enhancement cannot be due to an anomalous pollution of the pristine
 material from which these stars
were born, but that the s process must be 
at work in their interiors \cite{merr52}.
In addition, it was realized that only relatively faint AGB stars
become C-rich and that the brightest AGBs are N-rich \cite{wa98}. 
This evidence supports the hypothesis that C(N)-stars  have low 
mass progenitors. In massive AGBs, indeed, 
the CN-cycle taking place at the base of the
external convective layer converts most of the C dredged up into N.

\begin{figure}[h]
   \begin{center}
   \includegraphics[width=13cm]{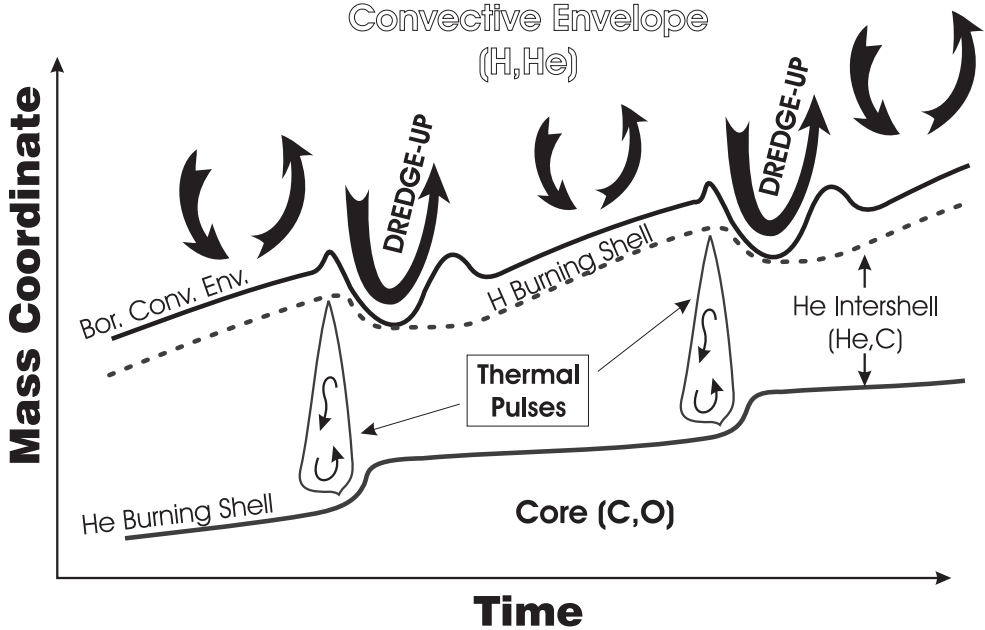}
   \end{center}
      \caption{This sketch illustrates the evolution of the positions 
      of the inner border of the convective envelope, the H-burning shell
       and the He-burning shell, during the thermally pulsing AGB phase.
       The convective regions generated by two subsequent
       thermal pulses are also shown. Note that the temporal developments of thermal
       pulses and of following TDU episodes are off scale with respect to the interpulse period
       (see Figure \ref{bordi} and \ref{marie}).}
         \label{cartoon}
   \end{figure}

Recent studies have demonstrated that the AGB stars with $M$ $<$ 3 $M_\odot$ are 
the major contributors to the galactic production of the 
isotopes belonging to the main component of the s process,
involving isotopes with atomic mass beyond $A \sim 80$
(see for a review \cite{wa97}).
Clayton \cite{cla67} first pointed out that a multiple 
neutron exposure, rather than a single exposure,
is needed to reproduce the main component and  
\cite{u73} suggested that this condition
could occur in the He-rich layer of an AGB star. 
A few years before, indeed,
\cite{scha67} discovered that during the late AGB
the He burning shell is recurrently switched on and off. 
At each He reignition, a thermonuclear 
runaway (thermal pulse) occurs and the rapid release of 
nuclear energy induces the formation of an extended convective region,
where the products of the nucleosynthesis are fully mixed. 
Following the original idea of \cite{u73}, 
the partial overlap of the convective regions recurrently generated by 
the thermal pulses ensures the required multiple neutron exposure (see Figure \ref{cartoon}).

Two promising  neutron sources for the build up of
the heavy s-process isotopes were early recognized 
(\cite{cam54}, \cite{cam60}): the 
\nean~ and the \ctan~ reactions. 
On the basis of detailed AGB stellar models,  
it has been understood that the major source of 
neutrons in low mass AGB 
stars is provided by the \ctan~ reaction.  
The s-process nucleosynthesis mostly occurs 
during the relatively long interpulse
period (namely the time elapsed between two subsequent 
thermal pulses), in a thin radiative layer at the top of 
the He intershell ($\sim$ 10$^{-3}$ $M_\odot$),
when the temperature ranges between 80 and 100 $\times$ 
$10^6$ K \cite{stra95}.
A second neutron burst giving rise to a small neutron 
exposure\footnote{We define the neutron exposure 
$\tau$ = $\int n_n v_{th} dt$ as 
the time integrated neutron flux, where  $n_n$ is the 
neutron density and $v_{th}$ the thermal velocity.}, but 
with a high peak neutron density, is released by the 
marginal activation of the $^{22}$Ne neutron source in 
the convective thermal pulse, modifying the final s-process 
composition at branchings along the s path depending on
the neutron density or on temperature.

In this paper we discuss the present status of the extant 
studies 
of nucleosynthesis and evolution in low mass AGB stars.
The AGB evolution is reviewed in sections \ref{sec2} and \ref{sec3}.
In section \ref{sectdu} the general problem of the third dredge up
efficiency in low mass AGBs of different mass and 
metallicity is discussed. An attempt to properly evaluate
the mass loss rate, a quantity that significantly affects the AGB evolution
 and nucleosynthesis, 
is presented in section \ref{secml}.
In section \ref{secnsou} the 
problem of the source of neutrons is afforded in some 
detail, with particular emphasis to the physical process driving the formation
of a $^{13}$C pocket in the He-rich and $^{12}$C-rich 
intershell.
An AGB stellar model of initial mass $M$ $=$ 2 \ms~ and solar 
metallicity, computed by coupling a stellar 
evolution code with a full nuclear network
(from H to Bi), is described in section \ref{resu}.
Finally, among the many aspects
related to the s-process nucleosynthesis occurring in 
AGB stars, we focus, in section \ref{seclead}, on one of the presently
most debated problems in the field, i.e.  
the expected s-process nucleosynthesis for the most
metal-poor stars in the Milky Way. 

\section{Approaching the Thermally Pulsing AGB}
\label{sec2}
In stars less massive than $\sim$ 10 $M_\odot$, central 
He burning leaves  
a compact C-O core. After He exhaustion, 
the central density rapidly increases 
(above $10^5$ and up to $10^8$ g/cm$^{-3}$), electrons become
highly degenerate
and a huge energy loss by plasma neutrinos takes place. The neutrino energy depletion
is only partially 
supplied by the release of gravitational energy from the core contraction, and 
the thermal content of the core is used to balance the deficit.
The resulting cooling starts from the center, where the density is higher,
and the maximum temperature moves progressively outward. 
It exists a maximum mass, called $M_{up}$, for which the 
whole core cools down, a fact that prevents carbon 
ignition. The precise value of $M_{up}$ depends 
on the chemical composition:  
it is $\sim$7 $M_\odot$ for population I stars (of nearly solar composition) and 
for very metal-poor stars (population III), while it is 
smaller ($\sim$6 $M_\odot$) for population II (halo) stars.   
Stars with mass slightly larger than $M_{\rm up}$ suffer a violent 
carbon ignition in degenerate condition 
(\cite{dti93}, \cite{rgi96}).
Stars with smaller mass enter the thermally pulsing AGB phase (see \cite{dom99} for an
updated presentation of the pre-AGB evolution).

\begin{figure}[h]
   \begin{center}
   \includegraphics[width=13cm]{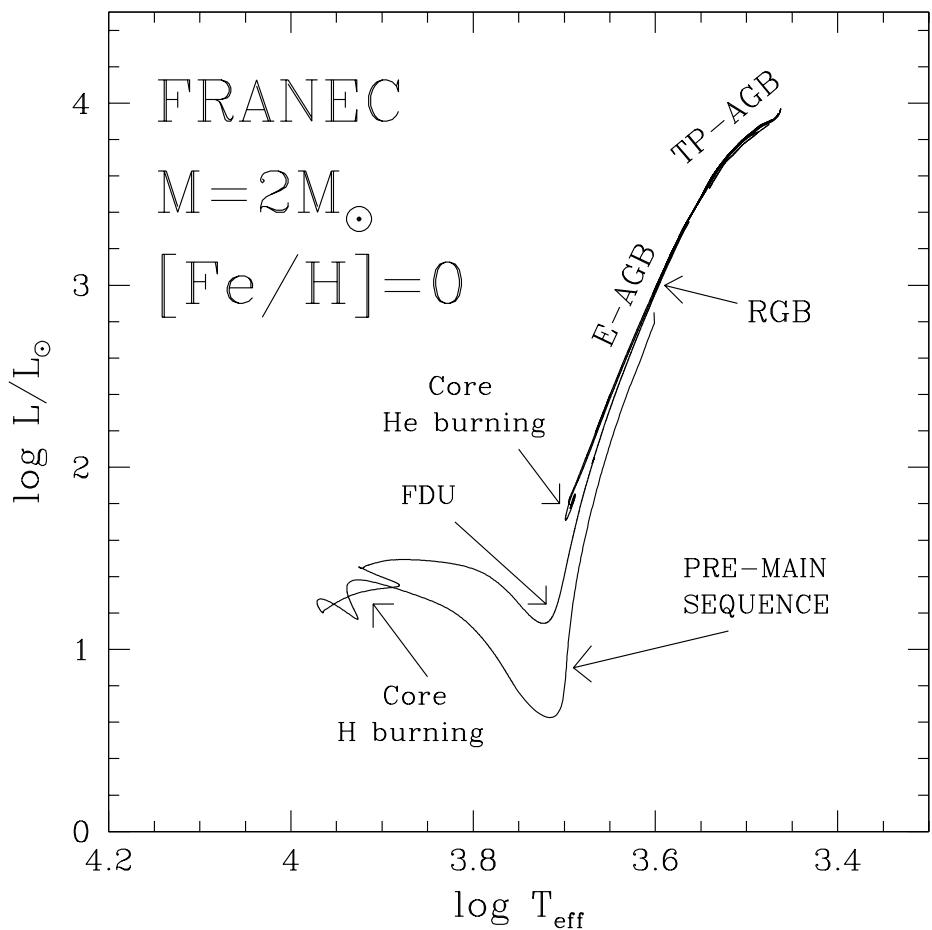}
   \end{center}
      \caption{Theoretical evolutionary track in the 
      Hertzsprung-Russell diagram of a star of 
        initial mass $M$ = 2 \ms~ and solar metallicity (see section \ref{resu} 
        for details of the model). }
         \label{fhr}
   \end{figure}

The evolutionary track in the HR diagram calculated for
a 2 $M_\odot$ star of solar metallicity is reported in
Figure \ref{fhr}. 
The computation has been started from a homogeneous
structure with a mass fraction of helium $Y$ = 0.275 and  
a metallicity $Z$ = 0.015\footnote{These values of $Y$ and $Z$ have been derived 
from a standard 
solar model, by fitting the luminosity ($L_\odot=3.844 \cdot 10^{33}$ erg/s), radius ($R_\odot=6.951 \cdot 10^{10}$ cm) and the mass fraction of metals relative to hydrogen ($(Z/X)_\odot=0.017$) 
of the present Sun.}.
The mass fractions of all the 
elements beyond hydrogen and helium relative to Z have been 
derived  from  \cite{ag89}, except for 
C and N \cite{ap02} and O, Ne and Ar \cite{asp04}.
The initial model is fully convective and corresponds to the pre-main 
sequence contraction phase, with a central temperature of   
about $10^5$ K. The model has been evolved from the
pre-main sequence, through core H burning, the red
giant branch (RGB), the off-center He flash and the He
burning,  up to the AGB.

After the star leaves the main sequence and first becomes a 
red giant, the 
convective envelope penetrates in the radiative
region above the H shell, enriching the surface
composition with the ashes of proton captures that occurred in
this deep zone during the main sequence (first dredge up, FDU);
with respect to the pristine composition,
$^4$He, $^3$He, $^{13}$C, $^{17}$O, and $^{14}$N are
enhanced, while $^{12}$C, $^{15}$N and $^{18}$O are
depleted (see e.g. \cite{bs99}).
Light elements, Li and Be, are practically extinguished.
The upper mass limit for the occurrence of the first dredge up
increases with the metallicity.
At $Z=0.02$, all stars entering the AGB have 
experienced the first dredge up episode, but this only occurs
for $M$ $<$ 3 $M_{\odot}$ and $M$ $<$ 2.5 $M_{\odot}$ when
$Z=0.001$ and $0.0001$, respectively (\cite{bs99}, \cite{dom99}).

During the first part of the asymptotic giant branch 
(called early-AGB), the He shell burning progressively moves outward and 
the mass of the C-O core increases.
In low mass stars, the H burning maintains an entropy barrier 
that limits the internal boundary of the external convective layer.
In contrast, in massive AGB stars, owing to the huge energy flux coming 
out from the He burning zone, the base of the H-rich envelope expands and cools, so that  
the H burning dies down. In this case,
the external convection penetrates inward, within the H-depleted zone. This is
the second dredge up (SDU), found in 
stellar models with $M$ $>$ 3$-$5 $M_{\odot}$, depending on 
the chemical composition\footnote{the lower the 
metallicity, the smaller the 
minimum mass for the occurrence of the second dredge up.}.
As a consequence of the SDU,
a further increase of the surface abundances of helium and
nitrogen is expected in massive AGBs. In addition, the SDU reduces the
H-depleted region and prevents the formation of massive white dwarfs. 

Finally, when the He burning shell gets closer
to the H/He discontinuity,   
it dies down and, after a rapid contraction,
the H burning shell fully supplies the surface energy loss.

\section{Thermally pulsing AGB}
\label{sec3}
\begin{figure}[h]
   \begin{center}
   \includegraphics[width=13cm]{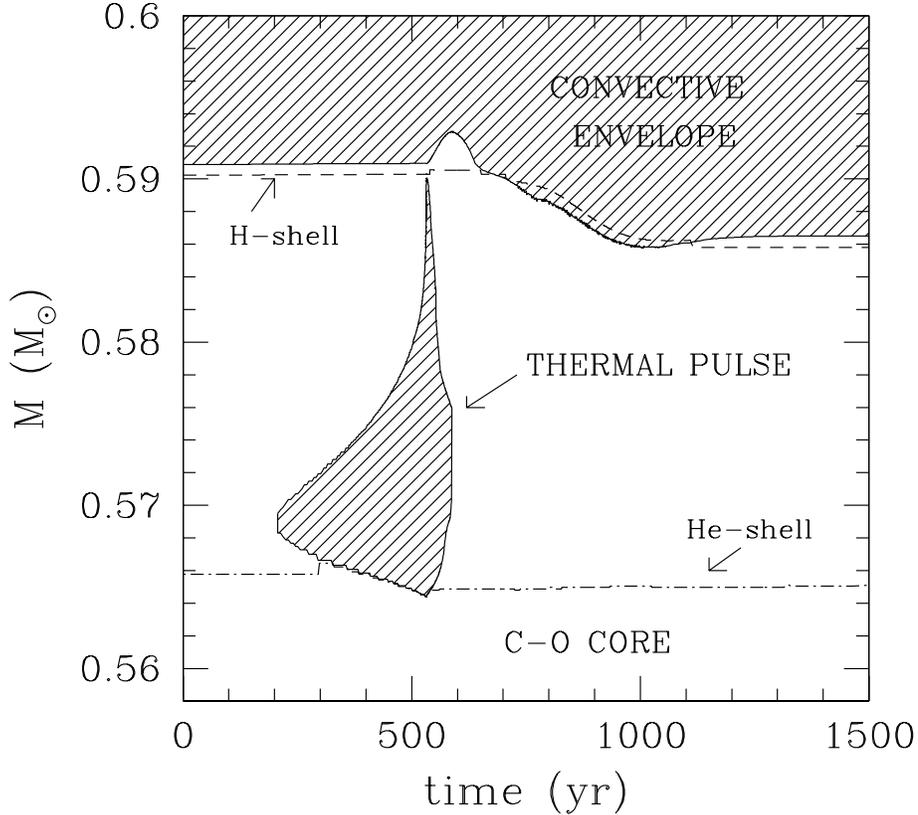}
   \end{center}
      \caption{Convective episode (dashed area) in the model 
      of $M$ = 2 $M_\odot$, $Z$ = 0.015 and $Y$ = 0.275, during and after the 10$^{\rm th}$ thermal pulse.
      Note how the convective zone generated by the thermal pulse covers the
      whole He intershell. After about 200 years, 
      the external convection penetrates inward (TDU).}
         \label{bordi}
   \end{figure}

\begin{figure}[h]
   \begin{center}
   \includegraphics[width=13cm]{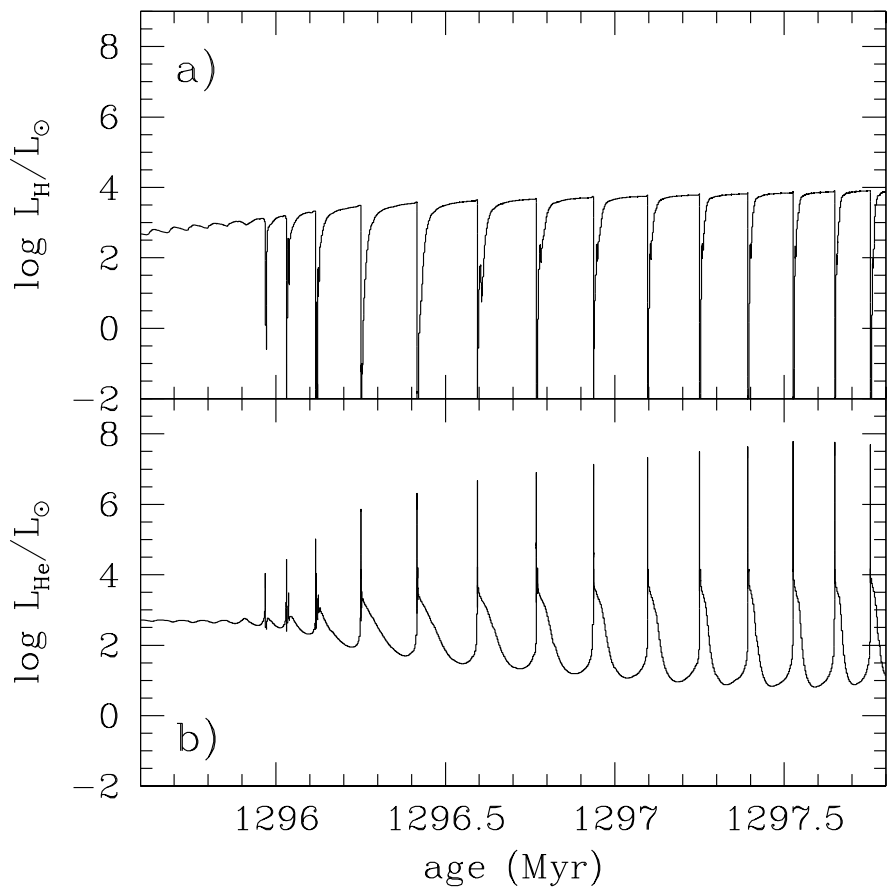}
   \end{center}
      \caption{Thermally pulsing AGB evolution of the model
      $M$ = 2 $M_\odot$, $Z$ = 0.015 and $Y$ = 0.275; 
      panel a): energy production rate of the H burning shell;
      panel b): energy production rate of the He burning shell.}
         \label{lumhhe}
   \end{figure}

The temporary stop of the He burning shell marks the 
beginning of the thermally pulsing AGB phase (TP-AGB).
The first thermal pulse occurs when the H burning shell accumulates
enough He below it, so that the He-rich zone is compressed and heated, and He reignites. 
Although the degree of electron degeneracy of the He-rich 
material is
weak, a thermonuclear runaway occurs, because the thermodynamical time scale
needed to locally expand 
the gas is much longer than the nuclear burning time scale of the
$3\alpha$ reaction \cite{scha65}. 
Owing to the fast release of nuclear energy, the local temperature
increases and the He burning luminosity blows up, in
extreme cases to $10^9$ $L_{\odot}$.
The thermonuclear runaway drives the formation of a 
convective zone 
that extends from the region of the partial He burning
to the H/He discontinuity (see Figure \ref{bordi}).
At the base of this convective shell an incomplete
He burning takes place and
the products of the $3\alpha$ reaction (essentially 
carbon) are mixed over the whole intershell.  At the quenching of a 
thermal instability, the
resulting mass fraction of C in the top layer  of the He 
intershell is $X$($^{12}$C) $\sim$ 0.25.
When the expansion 
has progressed far enough, the temperature of the He shell decreases
and a quiescent He burning phase begins.
The variations of the H and He burning luminosities for the 
2 $M_{\odot}$ model with $Z$ $=$ 0.015 are reported in Figure \ref{lumhhe}.
The expansion powered by the thermal instability pushes outward the layers of
material located above the He burning shell.
The temperature and the density at the base of the H-rich envelope
decrease and the H burning shell dies down. In these conditions, 
a third 
dredge up (TDU) episode is strongly favoured.
Indeed, at the interface between the envelope and the mantel, owing to the large 
energy flux coming from below, the local temperature gradient increases.
On the other hand, due to the low density, 
the ratio of the gas pressure to the radiation pressure decreases
and the adiabatic temperature gradient 
approaches its minimum allowed value for a fully ionized gas plus radiation 
(i.e. $(d$log$T/d$log$P)_{ad}$ $=$ 0.25).
Then the Schwarzschild criterion for the convective instability is more easily 
fulfilled and 
the envelope may penetrate (in mass) within the He intershell. 
The propagation of the convective instability is self-sustained due to the  
increase of the local opacity that occurs 
because fresh H (high opacity) is brought by convection into the He-rich 
layers (low opacity).
If the dredge up
is deep enough to overlap the region
previously mixed by the convective zone generated by the TP, 
helium, carbon and heavy s elements are brought to the surface. 
A C(N) star may eventually form as a consequence of a
series of recurrent TDU episodes 
(\cite{ib82}, \cite{lat89}, \cite{stra97}).

Between two subsequent pulses (interpulse phase), the 
H burning shell supplies the
energy radiated by the stellar surface and  
the luminosity of the star basically depends on the 
mass of the H-exhausted core, $M_{\rm H}$.
In other words, a direct correlation exists between the core mass 
and the stellar luminosity (\cite{pa70}, \cite{ir83})\footnote{The classical core mass-luminosity relation, 
as reported by \cite{ir83}, relates the maximum luminosity reached
toward the end of the interpulse phase and the mass of the H-exhausted core,
namely: $L_{max}=5.925\cdot10^4(M_H-0.495)$.}. 
As the H burning shell advances in mass, the internal 
boundary of the convective envelope recedes, but always remains 
very close to the region where the thermonuclear reactions 
are switched on.

If the stellar mass is sufficiently large
($M$ $\geq$ 5 $M_{\odot}$)\footnote{This lower limit is smaller at low metallicity.}, 
some nuclear processes are
indeed active within the most internal layers of 
the convective envelope.
In the more massive AGBs ($\sim$ 7 \ms), the temperature at the base of the 
convective envelope may reach 80 $\times$ $10^6$ K. This phenomenon is called
hot bottom burning (HBB) (\cite{su71}, 
\cite{ib73}, \cite{fc97}).
The main consequences of the coupling of nuclear burning 
and external convection are:
i) a substantial increase of the stellar luminosity, which deviates from the classical
core mass-luminosity relation (\cite{blsh}, \cite{bs92});
ii) a modification of the surface composition.
In particular, most of the C dredged up is converted into N, 
a fact that prevents the formation of massive
C(N)-stars. Moreover, the surface $^{12}$C$/^{13}$C ratio
approaches its equilibrium value (3.5) and an enhancement of Li is possible \cite{sb92}, \cite{vent}, 
through the so-called Cameron-Fowler mechanism \cite{cafo}.

\section{The third dredge up}
\label{sectdu}
The products of the s-process nucleosynthesis as well as those of the partial He burning
can be actually observed only if 
the third dredge up takes place. The TDU is driven by the expansion of the 
envelope powered by a thermal pulse. It is deeper 
when the strength of the pulse, as measured by the maximum
luminosity attained by the He burning, is stronger.
In principle, the strength of the pulse depends on temperature 
and density of the He-rich
layer at the epoch of the ignition.
The higher is the density and the temperature 
at He ignition,
the stronger the thermal pulse is. For this reason,
since He is accumulated by the thermonuclear fusion 
occurring in the H burning shell at the base of the 
envelope, the H burning rate is among the most important 
quantities regulating the physical conditions 
of the point where He ignites. 
It follows that the third dredge up is influenced
 by the parameters affecting the 
H-burning rate such as, in particular, the metallicity, the mass of the H-exhausted 
core and the mass
of the envelope (\cite{wo81}, \cite{ir83}, \cite{stra03}).
 As a general rule, a slower H burning 
implies a higher density of the
He-rich layer and, therefore, a stronger thermal pulse ensues.
Such an occurrence suggests a few important considerations:    

i) along the AGB, the mass of H-depleted material that is 
dredged up in a single episode  
($\delta$$M_{\rm TDU}$) initially increases, because the core mass increases, 
reaches a maximum and then
decreases, when the mass loss erodes a substantial fraction of the envelope
(see Figure \ref{tdutemp});

ii) for given core and envelope masses, the TDU is deeper in low metallicity stars,
 because the H burning
is less efficient. We found that  
$\delta$$M_{\rm TDU}$ ($Z$ = 0.002) 
$\sim$ 2 $\times$ $\delta$$M_{\rm TDU}$ ($Z$ = 0.02),
for the same core and envelope mass. The relation of the TDU 
with the metallicity
should be carefully considered when comparing disk AGB 
stars  with metal-poor extragalactic
AGB stars, like those belonging to the Magellanic Clouds or to the Dwarfs Spheroidals 
galaxies \cite{do04};

iii) there is a minimum envelope mass for which the TDU takes
 place. This minimum depends on the core mass and on the 
chemical composition of the envelope.
We found that the TDU ceases when the envelope mass becomes
smaller than 0.3 $\div$ 0.5 $M_\odot$ \cite{stra03}. 
This implies that in stars of initial mass below a given 
threshold 
($M$ $<$ 1.2 $M_\odot$), the residual envelope mass at the beginning of the
thermally pulsing AGB is already too small and the TDU cannot takes place.
As a matter of fact, AGB stars belonging to the 
Galactic Globular Clusters, whose initial mass
are of the order of 0.8 $-$ 0.9 $M_\odot$, do not show the enhancement of 
C and s-elements, which is the signature of the TDU
\cite{sek00}. Similarly, at variance with their
metal-rich disk counterparts, halo post-AGB  
stars do not show any significant enhancement of the s-elements \cite{glg00}.

\section{Mass loss}
\label{secml}
During the AGB, the star may become unstable 
against large amplitude pulsations. Pulsations induce
 a compression of the gas and the resulting increase of the density of the cool 
atmospheric layers favours the formation of complex molecules and dust grains, which
trap the outgoing radiation flux driving a strong wind.
Mass loss could also be influenced by the environment, as in  
close binary systems or in crowded stellar populations, 
like the central region of Globular Clusters. 
Mass loss erodes the envelope causing important changes 
in the stellar properties. The duration of the AGB, the strength of the pulse, the
efficiency of the third dredge up are a few examples of the quantities 
affected by the mass loss. The correct evaluation of the mass loss rate is 
also required to estimate the degree of chemical pollution of the
interstellar medium ascribed to AGB stars.

The AGB mass loss rate may be estimated from infrared 
color indices or from molecular CO rotational lines 
measurements. 
The available data indicate that it ranges between
$10^{-8}$ and $10^{-4}$ $M_{\odot}$/yr.
In variable AGB stars, the larger the period
the larger the mass loss is. 
No other correlations between mass loss and stellar parameters 
(luminosity, mass, composition)
have been clearly identified (see e.g. \cite{wh03}).  
The poor knowledge of the actual mass loss rate is one of the major 
uncertainties of AGB stellar models. 
Much effort has been devoted to derive a suitable prescription for this phenomenon
to be used in calculations of AGB evolutionary models. 
The Reimers' formula\footnote{
$\dot M (M_\odot/yr) = 1.34 \cdot 10^{-5} \cdot \eta \cdot \frac {L^{3 \over 2} } {M \cdot T_{\rm eff}^2}$, 
where $L$ and $M$ are in solar units and the effective temperature in K.} 
was introduced to describe the 
mass loss in population II red giants \cite{Rei75} and  
the $\eta$ parameter was calibrated according to the luminosity and color distribution of
bright globular clusters stars ($\eta \sim 0.4$, \cite{renzini81}). 
Unfortunately an equally 
stringent constraint for the calibration of the mass loss rate in AGB stars
is lacking.  
In principle, the mass loss rate may be adjusted in order to reproduce the observed 
luminosity functions
of the AGB stellar population,
or that of a sub-sample of AGBs, like the C(N)-stars. 
Another important constraint can be derived from the initial to final mass relation \cite{we00}. 
On the base of synthetic AGB models, \cite{grde93} suggested 
that a Reimers' mass loss formula,
with the parameter $\eta=5$, provides a suitable reproduction of these
observational constraints.
Their method, however, 
only indicates the average mass loss rate, whereas
the mass loss history remains largely unknown.
Indeed \cite{grde94} showed how  
different mass loss prescriptions may equally fulfil the same constraints.
Our tests, which make use of detailed stellar models to reproduce  
similar observational constraints, show that the mass
loss rate suggested
by \cite{grde93} may be adequate only for the more 
massive and/or more evolved AGB stars, 
but in low mass AGBs it is definitely too high.

\begin{figure}[htb]
   \begin{center}
   \includegraphics[width=13cm]{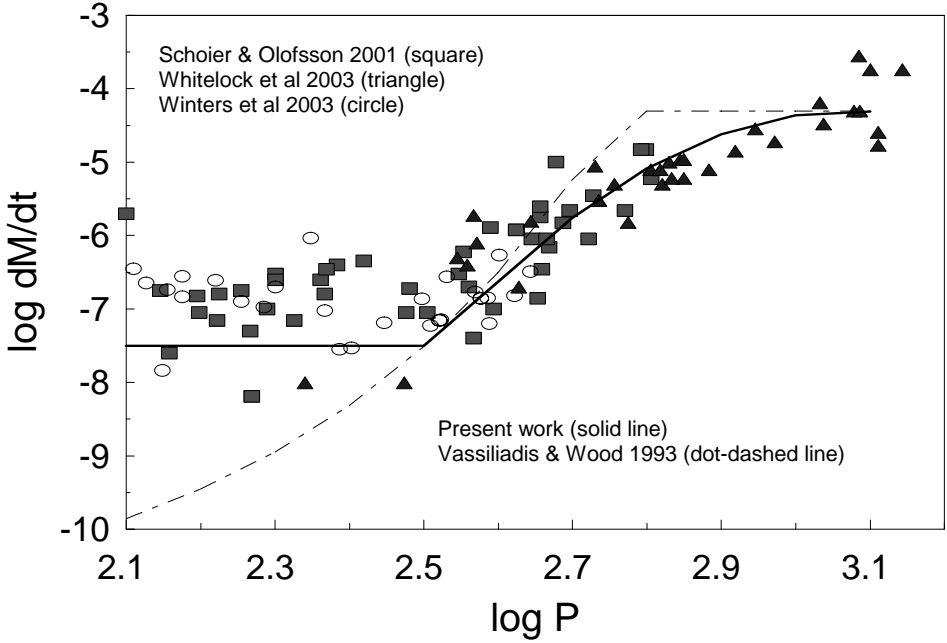}
   \end{center}
      \caption{Comparison of various mass loss rates versus period measurements
       (symbols) and
      prescriptions used in stellar evolution calculations (lines).}
         \label{ml1}
   \end{figure}

A possible alternative method to estimate the 
mass loss rate is based on the observed correlation with 
the pulsational period (see e.g. \cite{vw93}). 
Since the evolution of the pulsational 
period depends on the variations of radius, luminosity and mass,
this relation provides a simple 
method to estimate the evolution of the total stellar mass from basic stellar parameters.
In Figure \ref{ml1} we have collected data for the mass loss rate versus period, as measured by
various authors for O-rich (open circles) and C-rich (triangles and squares) AGB stars. 
Beside the evident spread of the data, 
three different regimes can be recognized: 
i) $P$ $<$ 300 days: moderate mass loss (roughly between $10^{-8}$ and 3 $\times$ $10^{-7}$ 
$M_\odot$/yr);  
ii) 300 $<$ $P$ $<$ 1000 days: exponential increase of the mass loss;
iii) $P$ $>$ 1000 days: the mass loss approaches a maximum ($\sim$ 5 $\times$ $10^{-5}$ 
$M_\odot$/yr), which roughly
coincides with the expected radiation-pressure-driven limit \cite{vw93}.
The dot-dashed line in the figure represents the prescription by Vassiliadis \& Wood \cite{vw93}. 
It clearly underestimates the mass loss rate of short period variables,
whilst for $P$ ranging between 500 and 1000 days the mass 
loss appears too high by an order of magnitude.
Thus, we have worked out a new relation (solid line), which provides 
a better calibration of the mass loss-period relation, namely:

for ~log$P < 2.5$ ~~~ log$\dot M = -7.7$

for ~$2.5 \le$log$P \le 3.1$ ~~~log$\dot M = -101.6 +63.26 \cdot$log$P -10.282 \cdot$(log$P)^2$

for ~log$P > 3.1$ ~~~ log$\dot M = -4.3$

\begin{figure}[h]
   \includegraphics[width=13cm]{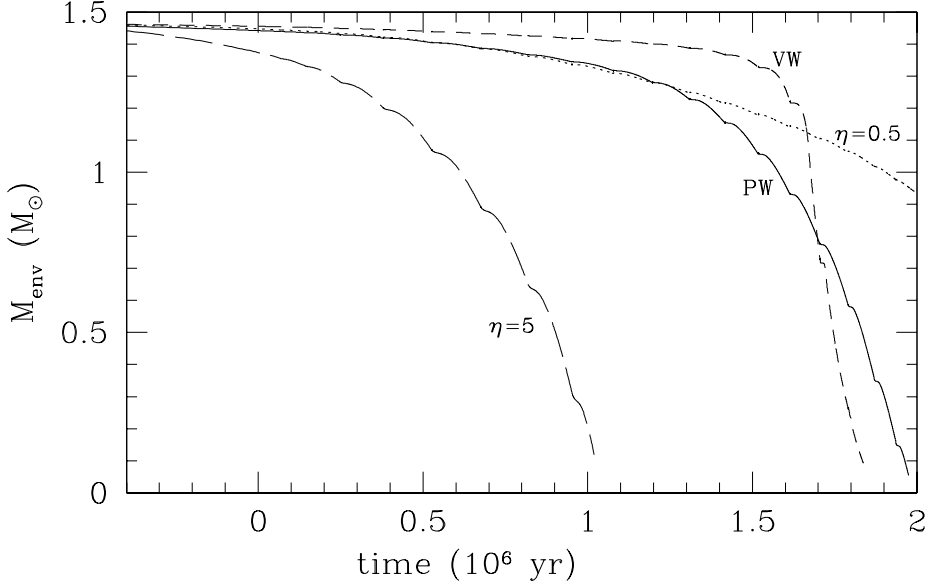}
      \caption{Variation versus time of the envelope
   mass according to different mass loss 
   prescriptions (see text), for an AGB model of initial mass $M$ = 2 \ms.}         
  \label{ml2}
   \end{figure}

This prescription has been used in our latest calculations of low mass 
AGB stellar models.
In Figure \ref{ml2} we report the evolution of the envelope mass ($M_{env}$)
of an AGB model of initial mass  2 $M_\odot$, as obtained under different 
prescriptions for the mass loss, namely: Reimers ($\eta$ = 5), Reimers ($\eta=0.5$),
Vassiliadis \& Wood (VW) and our new calibration of the mass loss-period relation (PW). The 0
of the temporal scale corresponds to the beginning of the TP-AGB phase.
Note how the mass loss history depends on the rate prescription.  
While the Reimers' formula provides a constant increase of the mass loss, 
the Vassiliadis and Wood rate
leads to a negligible mass loss for the major part of the AGB evolution, 
with a sudden
increase to the radiation-pressure-driven limit toward the end. Our prescription 
resembles a moderate Reimers ($\eta=0.5$) mass loss rate for the first 1.2 Myr 
and switches to the stronger  $\eta=5$ in the late TP-AGB phase.
Note how the different prescriptions affect the duration of the AGB and, in turn,
the AGB luminosity function and the 
estimated contribution of AGB stars to the Galactic chemical evolution. 

\section{The source of neutrons and the formation of the $^{13}$C pocket}
\label{secnsou}
As recalled in the Introduction, in thermally pulsing AGBs
two major neutron sources may operate 
within the intershell, driven by the \nean~ or by
the \ctan~reaction.

The amount of  $^{14}$N left by the H burning at the top of the intershell  
is practically equal
to the sum of the abundances (by number) of the C-N-O in the envelope. 
During the early phase of the convective thermal pulse, the material within the 
intershell is fully mixed and the
 $^{14}$N is totally converted into $^{22}$Ne throughout the chain
$^{14}$N($\alpha$,$\gamma$)$^{18}$F($\beta^+$$\nu$)$^{18}$O($\alpha$,$\gamma$)$^{22}$Ne.
Near the peak of the thermonuclear runaway, if the temperature is high enough
(i.e. $\sim$3.5 $\times$ 10$^8$ K),  
the $^{22}$Ne($\alpha$,n)$^{25}$Mg 
reaction may provide a significant neutron flux. 
Iben (\cite{i75a}) first 
demonstrated that this condition is fulfilled in 
intermediate mass stars. Actually,
in low mass stars ($M$ $<$ 3 $M_\odot$), the temperature 
at the 
base of the convective zone generated by the TP barely 
attains 3 $\times$ $10^8$ K and the $^{22}$Ne 
neutron source is marginally activated (\cite{it78}, 
\cite{stra97}, \cite{ga98}).

In low mass AGB stars, an alternative neutron source  
is provided by the reaction 
$^{13}$C($\alpha$,n)$^{16}$O. This reaction
requires a substantially lower temperature,
namely $\sim$90 $\times$ $10^6$ K, which is easily attained 
in the He intershell. Note that  between two subsequent 
thermal pulses, the H burning shell
leaves some $^{13}$C in the upper region of the 
He-rich intershell. The burning of this $^{13}$C, however,
produces a negligible neutron flux, even if, as a 
consequence of the TDU, 
the CNO in the envelope  may grow up to 10 times the solar
abundance. This occurs because, in the material processed 
by the CNO burning, the $^{14}$N is, in any case, two 
orders of magnitude more abundant than the $^{13}$C.
In practice, all the neutrons released by the $^{13}$C left 
by the H burning are captured by the abundant $^{14}$N. 
Then, an alternative source of $^{13}$C is needed, in a 
zone where $^{14}$N is depleted. 
The isotope $^{13}$C is produced via 
$^{12}$C(p,$\gamma$)$^{13}$N($\beta^+$$\nu$)$^{13}$C. After each thermal pulse, 
the material within the He-rich intershell is plenty 
of $^{12}$C and all the $^{14}$N has been converted 
into $^{22}$Ne, but it is also H depleted. So, the problem 
is how to inject a few protons into the
He-rich intershell.

Sackmann et al. \cite{sa74} suggested that the convective 
zone generated by the thermal pulse 
could extend beyond the H/He discontinuity.
In this case, protons are ingested downward, where the 
temperature is so high that the reaction chain 
$^{12}$C(p,$\gamma$)$^{13}$N($\beta^+$$\nu$)$^{13}$C($\alpha$,n)$^{16}$O 
would release neutrons. Against this hypothesis, 
\cite{ib76} argued that during the thermonuclear runaway the
H burning shell remains active and generates
an entropy barrier that prevents the penetration of the 
convective instability into the H-rich envelope. Extant AGB stellar
models show that only in very metal-poor stars, owing to 
the lack of CNO, ingestion of protons in the convective 
pulse may occur
(\cite{fu00}, \cite{iw04}, \cite{stra04}, \cite{suda}).
More favorable conditions for the formation of a 
$^{13}$C pocket within the intershell
are realized at the epoch of the TDU and during the 
so-called post-flash dip, the period
that immediately follows the TDU \cite{ir83}.
Indeed, as a consequence of the TDU, a sharp discontinuity between the
H-rich envelope and the He- and C-rich intershell forms. This condition 
is maintained until H burning is reignited. The
time elapsed from the maximum penetration of the 
convective envelope and H reignition
is about $10^4$ yr (for a star of 2 $M_\odot$).
During this period, different physical
mechanisms may contribute to the downward diffusion of a few
protons into the underlying radiative layer.
In such a case, at H reignition the top layers of 
the He intershell heat up
and a $^{13}$C pocket forms. Roughly speaking, the required
mixing process must be
able to diffuse about 10$^{-6}$ $M_{\odot}$ of H within a region as large as 
$\sim$ $10^{-3}$ $M_{\odot}$. Note that an excess of protons must be avoided, because
in that case the production of $^{13}$C is followed by the production of $^{14}$N 
(via the $^{13}$C$(p,\gamma)^{14}$N reaction). 
Straniero et al. \cite{stra95} demonstrated that the $^{13}$C formed in this way 
is fully consumed by the $^{13}$C($\alpha$,n)$^{16}$O
reaction in radiative conditions during the interpulse phase,
 when the temperature rises up to 
$\sim$ 90 $\times$ 10$^6$ K, giving rise to a large neutron exposure 
with a maximum neutron density of about 10$^7$ cm$^{-3}$. 
The expected s-process nucleosynthesis has been first calculated by \cite{ga98}.

Among the various attempts to find the possible mechanism responsible for the formation of the
 $^{13}$C pocket,
\cite{ib82} evaluated the timescale of the atomic diffusion driven by the 
sharp composition gradient left by the penetrating convective envelope at the time of the TDU and
concluded that this is a promising possibility to diffuse enough
protons from the envelope into the He-rich intershell. 
More recently \cite{he97}, inspired by 
hydrodynamical simulations of \cite{fr96},
invoked mechanical overshoot of material from the bottom of the convective
envelope into the underlying stable region. 
Langer et al. \cite{la99} calculated AGB models taking into account 
stellar rotation and 
found a certain mixing of protons into the intershell at the epoch
of the TDU.
However, this rotationally induced mixing does not stop when the convective 
envelope recedes, but continues during the interpulse period causing 
the contamination of the $^{13}$C pocket with too much $^{14}$N. 
Gravity waves have been investigated to explain the 
formation of the $^{13}$C pocket by \cite{dt03}, and 
\cite{bu04} show that a magnetic field of $10^8$ gauss,
comparable with the one
measured in white dwarfs, may induce a deep circulation below the 
convective envelope at the epoch of the TDU, 
allowing the required diffusion of protons
into the intershell.

Owing to  the lack of a reliable description of the 
physical phenomena that govern 
the diffusion of protons into the He intershell,
 in the current AGB nucleosynthesis
calculations the amount of $^{13}$C is assumed as free parameter
 (see \cite{ga98}). In addition, the simultaneous solution of the
 stellar structure equations and a full network including all the relevant
 isotopes up to the termination point of the s-process path (Pb-Bi) requires a relevant 
 computational power. For this reason, a post-process nucleosynthesis calculation,
  based on AGB stellar models computed with
 a restricted nuclear network, is generally preferred. 
 In spite of these limitations, this approach provided
 a substantial improvement with respect to prior 
 calculations, based on the so-called
 classical analysis of the s process (see discussion in \cite{ar99}). 
 The adequacy of detailed post-process calculations has been proved by
 comparing theoretical expectations with observed abundances found in galactic disk AGB stars
 (\cite{la95}, \cite{ab01}, \cite{ab02}, \cite{re04}, \cite{bu00}).
 Additional confirmations have been derived from presolar SiC 
 grains found in meteorites
 (\cite{lu03}, \cite{sav04}, \cite{zi99}).
These calculations are the bricks of the extant chemical evolution models 
incorporating heavy elements \cite{tr04}.
In general, 
observations show an important scatter of  s-process abundances at any metallicity 
\cite{bu01}.
This scatter reveals a spread of the effective amount of 
$^{13}$C in the pocket, perhaps due to differences in the stellar parameters, 
such as the initial mass or the mass loss rate, or to the chaotic nature of the
process responsible of the proton injection into the He intershell.

\section {The nuclear network}
The next step in the understanding of the s-process 
nucleosynthesis in AGB stars is 
the development of algorithms  that allow the 
formation of the 
$^{13}$C pocket and a full coupling of the equations describing 
the physical and the 
chemical evolution of the star. In this framework,
we have included a full network of about 500 isotopes
(from H to the Pb-Bi-Po ending point) linked by more than 750 reactions, 
in the stellar evolution code
FRANEC \cite{cls98}. This network, the same already used in the post-process calculations,
is continuously upgraded according to 
the latest theoretical and experimental nuclear physics improvements.  
The adopted nuclear reactions are summarized in the following five subsections.

\subsection{Charged particle capture reactions}
Reaction rates of isotopes involving charged particles are generally taken
from the NACRE compilation \cite{Angulo}. For those reactions not included in this
database, we use the rates tabulated by \cite{cf88}.
Two important exceptions should be recalled.
Concerning \nqpg, we use 
the result of the recent measurement obtained by the LUNA collaboration \cite{luna}.
Being the bottleneck
of the CNO cycle,
this reaction regulates the He deposition on the intershell of an AGB star and in turn 
affects the pulse strength and the efficiency of the TDU (see the discussion in section \ref{sectdu}).
The new low energy astrophysical factor ($S$(0)=1.7 MeV barn) is about 40\% lower than that
suggested by NACRE.
With regard to $^{12}$C$(\alpha,\gamma)^{16}$O we adopt the rate recently derived
 by \cite{kunz}).

\subsection{Neutron sources}
\label{neusc}
 A subthreshold resonance makes difficult the low energy extrapolation of the
 $^{13}$C($\alpha$,n)$^{16}$O reaction rate.
We adopt the rate suggested by \cite{drot}. We have also investigated the effects
of different rates (\cite{kubo}, \cite{Angulo}) and we found a negligible 
modification of the final surface isotopic distribution (\cite{crinic},
see also, \cite{pigna}).
The $^{22}$Ne($\alpha$,n)$^{25}$Mg rate is even more 
complex,
 owing to the possible existence of unknown low energy resonances
\cite{koene22}. We adopted the lower limit suggested by \cite{kaep}. It
does not include the possible resonance at 633 keV and adopt a 1 $\sigma$ 
lower limit (164 $\mu$eV) for the strength of the resonance at 828 keV. 
The recommended value by the NACRE compilation at $T$ $\sim$ 
3 $\times$ 10$^8$ K is a factor of two lower than our
adopted choice, and very close to the recent determination 
of \cite{jaeg01}. With such a rate, only 
a few neutron-rich isotopes involved in branchings sensitive 
to the peak neutron density, like $^{86}$Kr, $^{87}$Rb,
 $^{96}$Zr, would be affected.
 
\subsection{{\it n}-capture reactions}
With {\it n}-capture reactions we mean ($n,\gamma$), ($n,\alpha$) and ($n,p$) processes.
For the ($n,\gamma$) reactions, we adopt,  as reference compilation,
 the recommended rates by 
\cite{bao} (hereafter BK2000). In that paper, experimental and theoretical
data are critically revised and the reaction rates are listed as a function of the 
thermal energy from 5 to 100 keV. We recall that the typical thermal energy at which
the s-process nucleosynthesis occurs in AGB stars are 8 keV for the 
$^{13}$C($\alpha$,n)$^{16}$O reaction  and 23 keV for the 
$^{22}$Ne($\alpha$,n)$^{25}$Mg reaction.
For the few reactions not included in BK2000, we use 
the theoretical calculations of \cite {rt2000}.

Starting from this database, we have upgraded the network with the most recent 
experimental results. With regard to Si isotopes we refer
to \cite{Guber2003}, Cl isotopes to \cite{Guber2002}, $^{60}$Ni to \cite{corvi}, $^{62}$Ni to \cite{RauscNi}, 
$^{88}$Sr to \cite{KoehSr88}, Kr isotopes to \cite{mutti}, Xe isotopes to \cite{reifhXe}, Cd isotopes to \cite{wiss}, 
$^{139}$La to \cite{obrien}, Pm isotopes to \cite{reifhPm}, $^{151}$Sm and Eu isotopes to \cite{best}
and finally Pt isotopes to \cite{KoehPt}.
The rates of ($n,\alpha$) and ($n,p$) reactions involving heavy isotopes are taken from RT2000. 
We adopted  the ($n,\alpha$) and ($n,p$) rates 
on light isotopes from various authors.
In particular: the \nqnp~ is taken from  \cite{Koehob}, 
the \odna~ from \cite{wageO17}, the \alnp~ from \cite{Koehal}, the \alna~ from \cite{wageal} and the \stna~ 
from \cite{schatz}.
 The \ctnp~ rate has been derived  from \cite{druy}, the \csnp~ and the \csna~ 
from \cite{wagecl}, the \asnp~ and the \asna~ from \cite{goemar}. The \anna~ is from \cite{goephd},
while \canp~ and \cana~ are from \cite{wageca}.

\subsection{$\beta$ decay reactions}
Weak interaction rates 
(electron captures, $\beta$ and positron decays) are interpolated as 
a function of the temperature (from $10^7$~K to $10^{10}$~K) and electron density
 (from 1 to 30 cm$^{-3}$). 
At temperatures lower than $10^7$~K we assume a constant value equal to the terrestrial one.
For isotopes up to $^{37}$Ar, data have been taken from \cite{oda}, 
with the exception of $^{7}$Be \cite{cf88} and 
the isomeric state of $^{26}$Al, for which we refer to \cite{coc}.
Concerning the unstable isotopes between $^{39}$Ar and $^{45}$Ca we use
 prescriptions by \cite{ffn8}, while  between $^{45}$Ca and $^{64}$Cu (excluding $^{63}$Ni)
we follow \cite{lp00}. 
For $^{63}$Ni and heavier isotopes we use the rates tabulated in \cite{taka}, with the exceptions 
of $^{79}$Se and $^{176}$Lu, for which we refer respectively to \cite{klay1} and \cite{klay2}.
For the few rates not included in any compilation, we use the terrestrial value.

\subsection{Isomeric states}
The existence of isomeric states of unthermalized isotopes leads to ramifications of the s-process
flux (\cite{wnc76}, \cite{wn78}). In particular,  branching points originated by the isomeric state of $^{26}$Al, 
$^{85}$Kr, $^{176}$Lu and $^{180}$Ta require particular attention.

Concerning $^{26}$Al, the proton capture on $^{25}$Mg 
is split in two distinct reactions: 
the first produces the ground state $^{26}$Al$^g$ (with terrestrial half life
$T_{1/2}$ = 7.16 $\times$ 10$^5$ yr) and the second creates the isomer $^{26}$Al$^m$ 
that almost istantaneously decays into $^{26}$Mg.

Concerning $^{85}$Kr, the neutron capture of $^{84}$Kr to $^{85}$Kr$^m$ 
has a 50\% probability with respect the total cross section (at 30 keV, \cite{be91}). 
$^{85}$Kr$^m$ has a non-zero probability to decay by internal conversion 
to its ground state (of 20\%), 
thus leading to the following isomeric ratio (IR):

\begin{equation} 
IR=\frac{\sigma(^{84}{\rm Kr}(n,\gamma)^{85}{\rm Kr}^m)}{\sigma_{tot}(^{84}{\rm Kr}(n,\gamma)^{85}{\rm Kr})}=0.42~.
\end{equation}

Note that the $\beta^-$ half life of $^{85}$Kr$^m$ is 4.48 h, while that of $^{85}$Kr$^g$ is 10.76 yr.

Concerning $^{175}{\rm Lu}$, below 20 keV we adopt the terrestrial isomeric ratio, namely:

\begin{equation} 
IR=\frac{\sigma(^{175}{\rm Lu}(n,\gamma)^{176}{\rm Lu}^m)}{\sigma_{tot}(^{175}{\rm Lu}(n,\gamma)^{176}{\rm Lu})}=0.11,
\end{equation}

while above this energy, the partial thermalization through a mediating state 
should also be considered (see \cite{klay2})
and we assumed an effective IR=0.5.

Finally, for the complex treatment of the branching between isomeric and ground state
of $^{180}$Ta, we refer to \cite{neme}.

\section{Mixing during the third dredge up and formation of the
$^{13}${\rm C} pocket}
In the model presented here we adopt 
a simple approach based on an argument early discussed by \cite{bi79}
(see also \cite{ccs92}, \cite{frl96}).
When the TDU takes place, the
opacity of the envelope (H-rich) is significantly larger than 
the opacity of the underlying H-exhausted region (He-rich). This fact causes an abrupt change
of the temperature gradient at the inner border of the penetrating convective envelope.
 In this condition, 
the convective boundary becomes unstable, because
any perturbation causing an excess of mixing
immediately leads to an increase of the opacity and, in turn, to an increase of the 
temperature gradient. This occurrence favours a deeper penetration of the convective 
instability or, in other words, a deeper dredge up. 
A similar mechanism is responsible for the growth of the convective
 core during central He burning (\cite{pa70}, \cite{cgr71}). 
A different view of the same phenomenon concerns the evaluation of the average convective velocity.
In the framework of the mixing length theory,
this velocity is proportional to the difference between the 
radiative temperature gradient (i.e. the gradient necessary to 
carry out the total energy flux if convection would be inhibited)
 and the adiabatic temperature gradient. For this reason, the average convective velocity
usually drops to 0 at the stable boundary of a convective
layer, where the temperature gradient coincides with the adiabatic one.
 However, when convection penetrates in a region of lower opacity (as 
happens during a third
dredge up episode), 
the difference between the actual temperature gradient and the adiabatic gradient 
grows above 0 and a positive average convective velocity is found at the 
inner border of the convective envelope, which, for this reason, becomes unstable.  
In principle, as soon as He is mixed with the envelope, the
opacity and, in turn, the difference between the radiative and the 
adiabatic temperature gradients are reduced. However, the 
mass of the convective envelope is usually much larger than the 
amount of the dredged up material
and the relaxing effect of the additional mixing is, 
in practice, negligible (\cite{frl96}, \cite{cms98}).
Then, a simple thermodynamic criterion cannot be used to determine the real extension of the
convective instability. For sure, the steep pressure gradient that develops immediately
below the formal border of the convective envelope limits the penetration of the instability,
so that the average convective velocity should rapidly drop to 0.
In order to mimic this behaviour, we assume that in the region underlying the formal 
convective boundary, the average velocity follows an exponential decline, namely:

\begin{equation} \label{param}
v=v_{bce}\exp{
\left(
-\frac{d}{\beta H_P}
\right)
} \;     ,
\end{equation}

where {\it d} is the distance from the formal convective boundary,
$v_{bce}$ is the velocity of the most internal convective mesh,
$H_P$ is the pressure scale height and $\beta$ is a free parameter.
Note that this formula is similar to the "overshoot" proposed by \cite{he97}. However, 
in our case, since $v_{bce}$ is usually 0,
it produces a negligible amount of extra mixing, except during a dredge up.

\begin{figure}[h]
   \begin{center}
   \includegraphics[width=13cm]{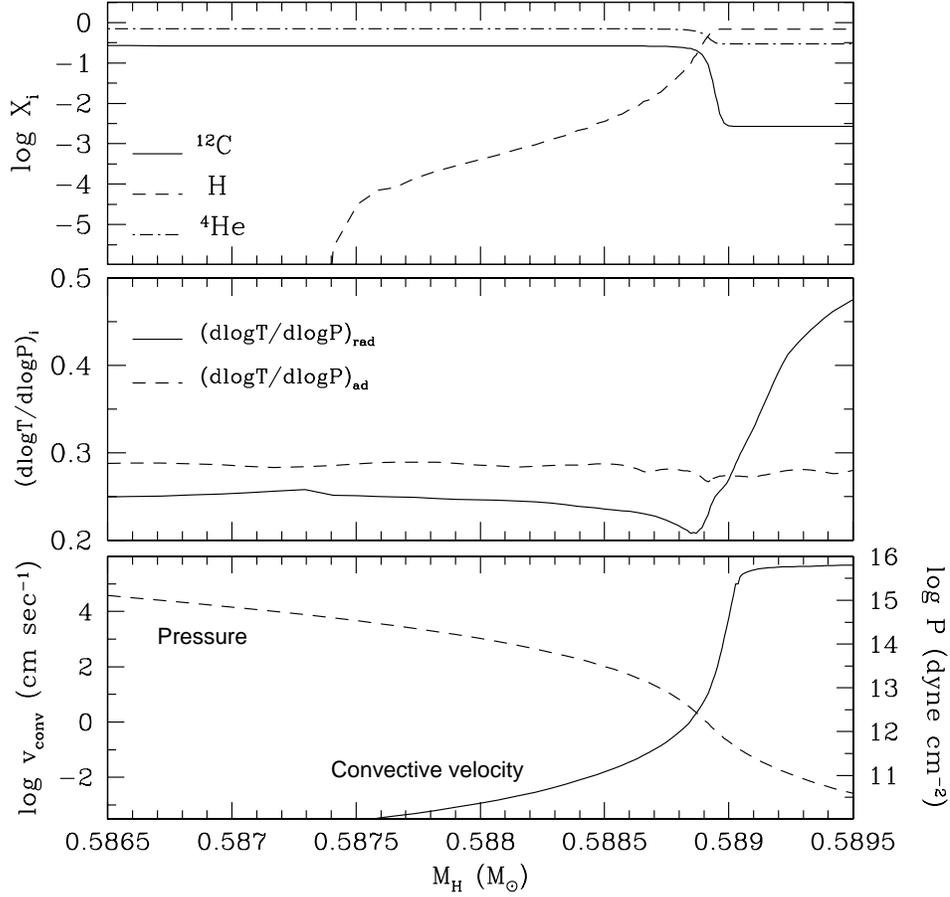}
   \end{center}
      \caption{Chemical profile ({\it upper panel}), temperature gradients ({\it middle panel}), average
        velocity and pressure ({\it lower panel}), in the region 
        at the convective boundary during the fifth TDU episode.}
         \label{velco}
   \end{figure}

Then, the degree of mixing is calculated by means of the following relation:

\begin{equation}
X_j=X_j^o+\frac{1}{M_{conv}}{\sum}_{k}(X^o_k-X^o_j)f_{j,k}{\Delta}M_k
\end{equation}
where the summation is extended over the whole convective zone and the
superscript $^o$ refers to unmixed abundances. ${\Delta}M_k$ is the mass of the mesh-point $k$ 
and $M_{conv}$ is the total mass of the convective zone.
The damping factor $f$ is:
\begin{equation}
f_{j,k}=\frac{{\Delta}t}{{\tau}_{j,k}}    
\end{equation}
if ${\Delta}t<{\tau}_{j,k}$,  or
\begin{equation}
f_{j,k}=1                  
\end{equation}  
if ${\Delta}t\ge{\tau}_{j,k}$.
Here ${\Delta}t$ is the time step and $\tau_{j,k}$ is the mixing turnover time
between the mesh-points {\it j} and {\it k}, namely: 
\begin{equation}
\tau_{j,k}=\int_{r(j)}^{r(k)} \frac{dr}{v(r)}={\sum}_{i=j,k}\frac{{\Delta}r_i}{v_i}   
\end{equation}
The mixing velocity ($v_i$) is computed according to the mixing length theory, except 
in the region where the exponential decline is assumed.
This algorithm allows us to account for the partial 
mixing that occurs when the time step 
is reduced to or below the mixing timescale.
In practice, when the third dredge up takes place, 
complete mixing is obtained within the fully convective zone,
while the region immediately below, where the convective turnover time scale is larger,
is only partially mixed.
Figure \ref{velco} shows the chemical profile, the radiative and adiabatic temperature gradients, the average
velocity and the pressure, in the region around the 
convective boundary, in a model of 2
$M_\odot$, during the 5$^{th}$ dredge up episode (see section \ref{resu}). 
The three major effects of the introduction of this exponential decay of the
convective velocity are:
i) the convective boundary is more stable against 
perturbation;
ii) a smooth profile of protons within the intershell is 
left by the TDU; iii) a more efficient TDU results.
As a consequence of ii), a $^{13}$C pocket almost $^{14}$N-free forms in this zone.
After a few tests, we found that in order to get a suitable amount of $^{13}$C
in the pocket,  
the $\beta$ parameter should be of the order of 0.1. 
\begin{figure}[h]
   \begin{center}
   \includegraphics[width=13cm]{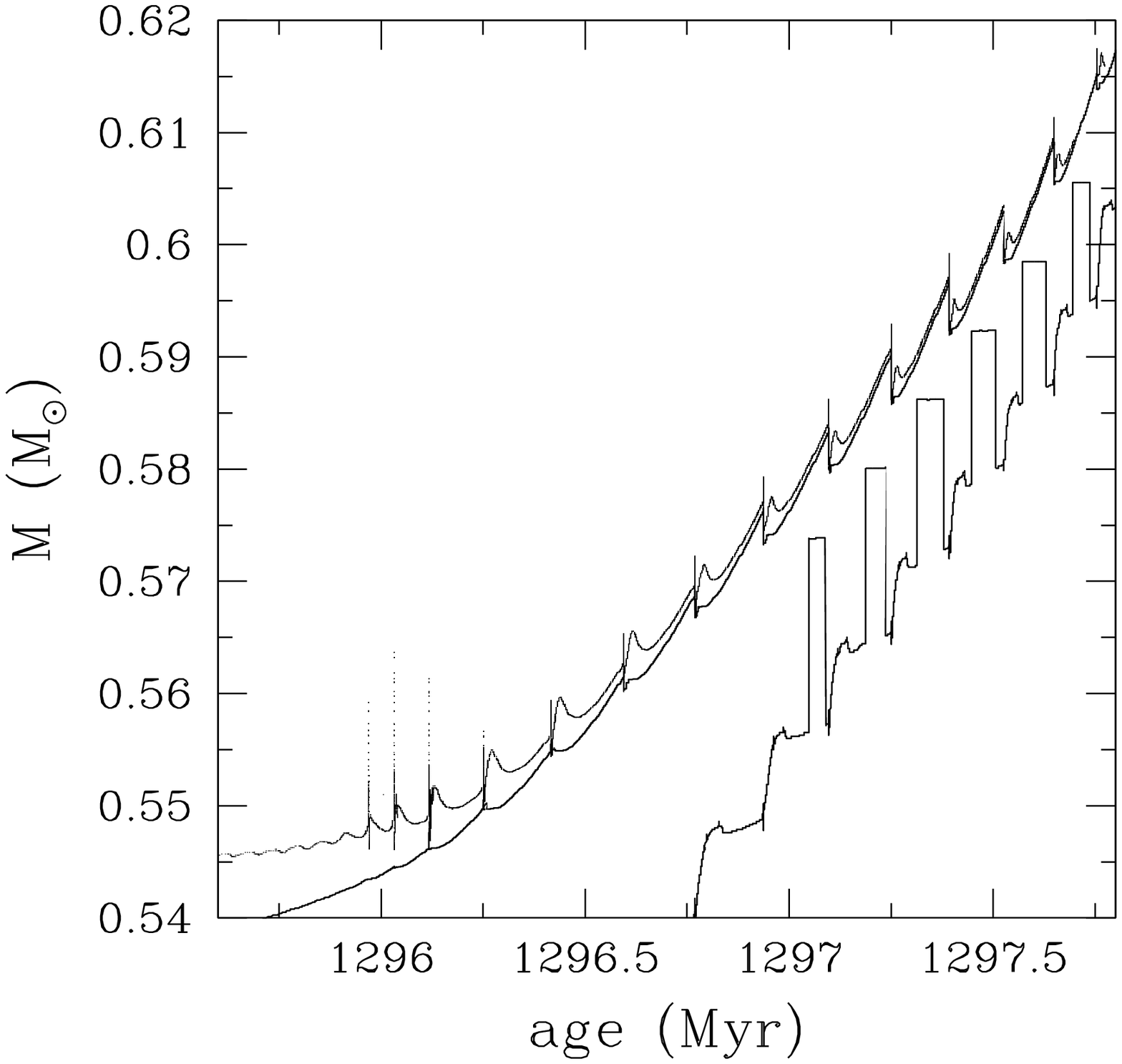}
   \end{center}
      \caption{The evolution throughout the TP-AGB phase of a star with 
  initial mass $M$ $=$ 2 \ms~ and solar metallicity.
  The three lines represent (from the top): the inner border of the convective envelope,
  the location of the maximum energy production within the H burning shell and the
  location of the maximum energy production within the He-rich intershell.}
         \label{marie}
   \end{figure}

\section {A thermally pulsing AGB model} 
\label{resu}
The network and the mixing algorithm described in the 
previous sections 
have been used to calculate the evolutionary sequence 
of a 2 $M_\odot$ star of solar metallicity shown in Figure 
\ref{fhr}.
Before the first thermal pulse, a Reimers' formula for the 
mass loss ($\eta=0.4$) has been assumed.
At this epoch, the total and the core masses are 1.95 
and 0.55 $M_\odot$, respectively.
During the thermally pulsing AGB phase, we use the 
mass loss rate prescription discussed in section 2.3
(see the solid line in Figure \ref{ml2}). The calculation 
has been stopped when the envelope mass
was reduced down to $\sim$ 0.3 $M_\odot$, the 
corresponding total mass being
 0.92 $M_\odot$. At this epoch, the TDU ceases (see Figure \ref{marie}) and hence the envelope composition no longer changes. 
Figure \ref{marie} illustrates the evolution of the mass 
coordinates of the maximum energy production 
within the He intershell, the maximum energy production 
within the H burning shell
and the inner border of the convective envelope. 
\begin{figure}[h]
   \begin{center}
   \includegraphics[width=13cm]{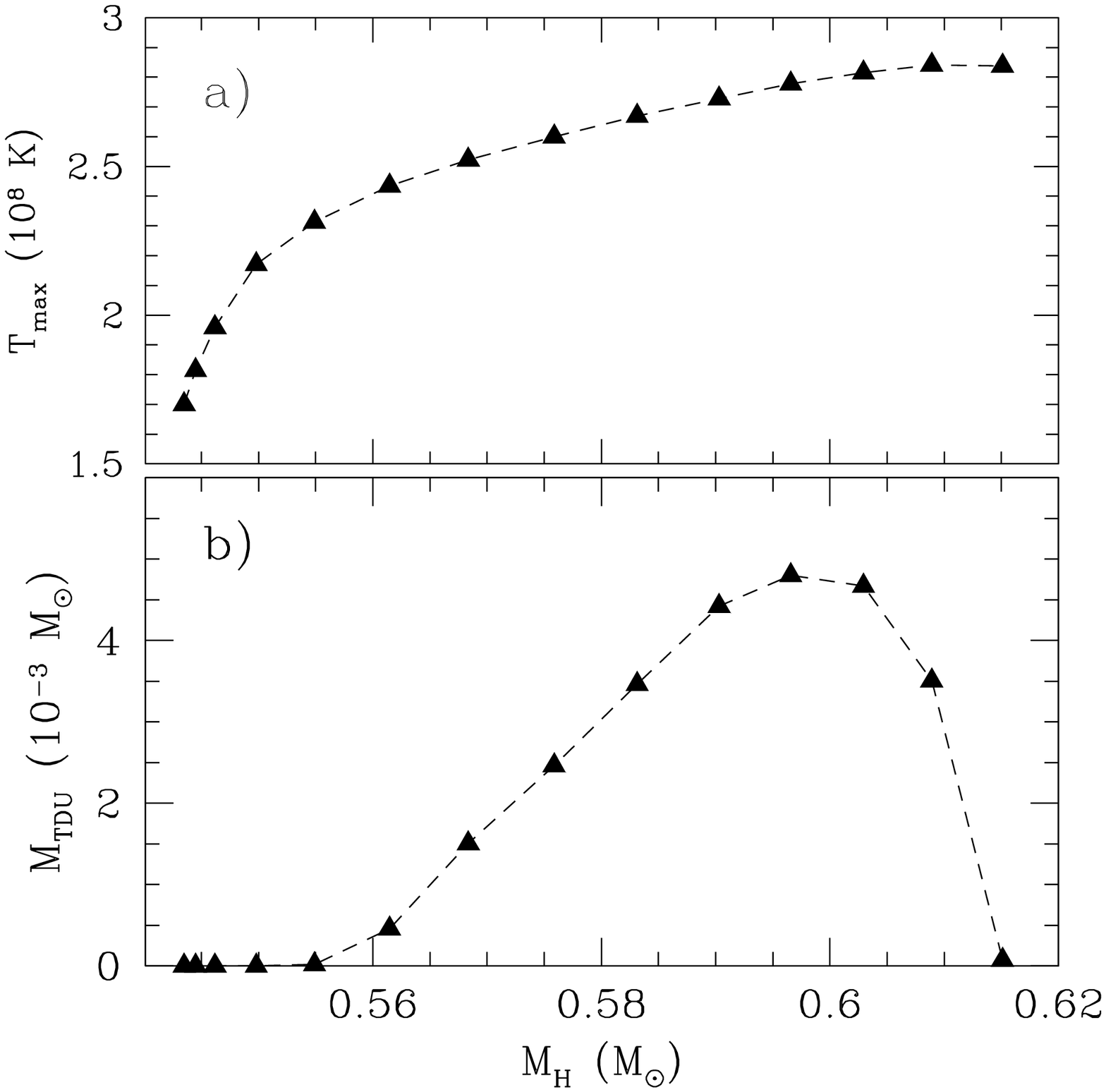}
   \end{center}
      \caption{Evolution of the maximum temperature at the 
base of the convective zone generated by the thermal pulse (panel a) and  
      dredged up mass per pulse (panel b) during the AGB 
phase of the model of Figure 1.}
         \label{tdutemp}
   \end{figure}
The first TDU episode occurs after six TPs, when the core 
mass is 0.56 $M_\odot$.
The evolution of $\delta$$M_{\rm TDU}$ (the amount of 
mass dredged up) and $T_{max}^{\rm TP}$
(the maximum temperature at the base of the convective 
zone generated by the thermal pulse)
are reported in Figure \ref{tdutemp}.
The possible activation of the \nean~ reaction critically depends
on $T_{max}^{\rm TP}$ that, for this
 reason, represents a key quantity for the comprehension of the 
nucleosynthesis occurring in AGB stars. It is   
obviously related to the maximum He burning luminosity.
Then, as for the dredge up, 
the maximum temperature is a function of the core mass,
the envelope mass and the composition \cite{stra03}.
As expected, the deepness of the third dredge up initially 
increases, due to the increase of the core mass. 
Then, during the TP-AGB phase, the effect of the envelope erosion
becomes important, and the dredge up efficiency decreases, dropping to 0 
after the last computed thermal pulse. 
C/O becomes larger than 1 when the core mass is about 
0.6 $M_\odot$ and the 
luminosity is log $L/L_\odot$ = 3.85, which corresponds to 
a bolometric magnitude of $-$4.9 mag. 
We recall that the majority of the C(N) stars in the 
galactic disk have bolometric magnitudes around $-$5 mag. 
The final C/O is 1.67.

\begin{figure}[h]
   \begin{center}
   \includegraphics[width=13cm]{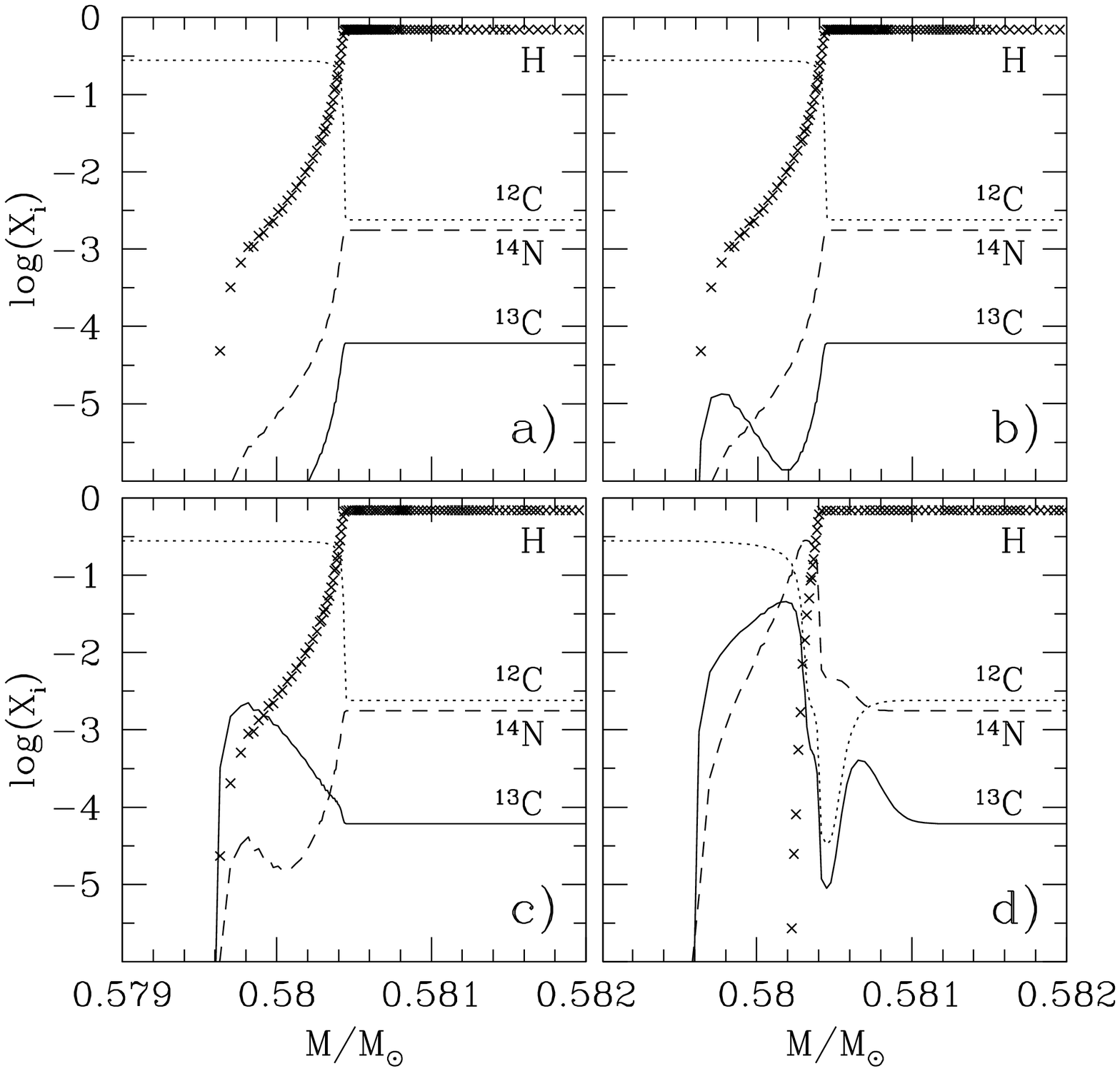}
   \end{center}
      \caption{The variation of the relevant chemical species in the region where the \ct~pocket 
      forms after the occurrence of the third dredge up. Each panel refers to different time steps, namely: 
      $\Delta t$ = 78, 1000, 2000 and 14000 years, respectively.}
         \label{evo}
\end{figure}

$T_{max}^{\rm TP}$ initially increases 
 and attains a nearly asymptotic upper limit, of
about 285 $\times$ $10^6$ K. 
At such a low temperature the  $^{22}$Ne($\alpha$,n)$^{25}$Mg is marginally activated.
On the contrary, an important s-process nucleosynthesis is driven by the 
$^{13}$C($\alpha$,n)$^{16}$O reaction. 
A $^{13}$C pocket forms after each TDU episode. 
The $^{13}$C in the first two pockets is only partially burned during the interpulse
and the residual
is engulfed into the  convective zone generated by the subsequent thermal pulse.
In all other cases, the $^{13}$C is fully consumed during the interpulse.
The outward shift of the location of the maximum energy
production within the intershell, in Figure \ref{marie},
marks the onset of the $^{13}$C burning during the interpulse.
Figure \ref{evo} illustrates the various steps of the formation of the $4^{th}$ $^{13}$C pocket.
Note that the 
zone where the $^{13}$C pocket forms (panel d)) is partially overlapped by 
a more external thin $^{14}$N pocket.
The maximum neutron density (about $10^7$ cm$^{-3}$) is attained in the more internal layer of the $^{13}$C pocket, where the $^{14}$N is less abundant.
\begin{figure}[h]
   \begin{center}
   \includegraphics[width=8cm]{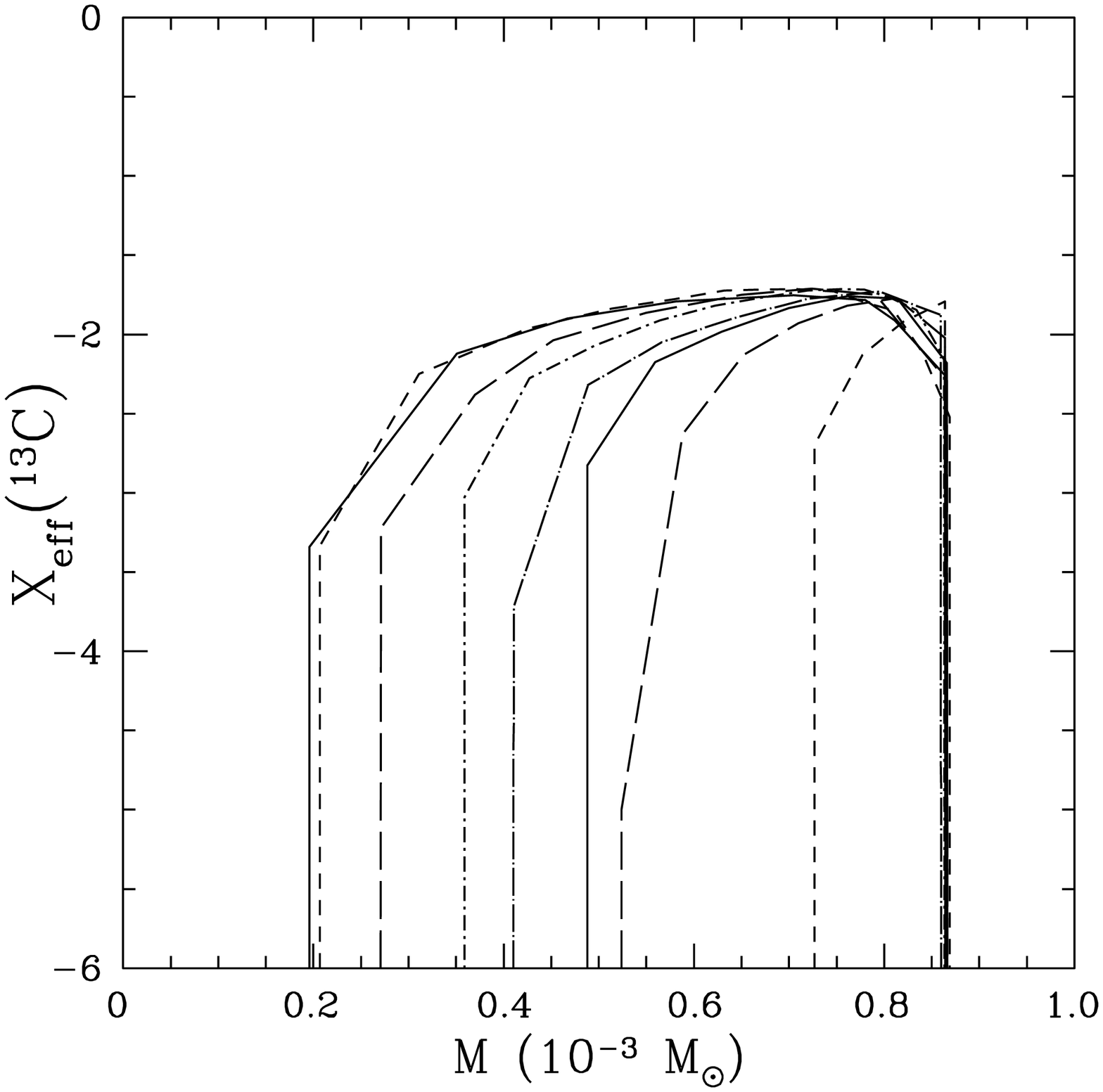}
   \end{center}
      \caption{The {\it effective} (see text) mass fraction of $^{13}$C 
within all
the pockets is reported. Each pocket has been shifted in mass in 
order to superimpose their external borders. 
The 0 point of the abscissa is arbitrary. The extension of the pocket
decreases with time. The first is about $6.5 \cdot 10^{-4}$ $M_\odot$ large.
The last is about 6 time smaller.}
         \label{evta}
\end{figure}
The {\it effective} abundances of $^{13}$C in all the pockets we found (defined as 
$X_{\rm eff}$($^{13}$C)=$X$($^{13}$C) - $X$($^{14}$N) $\times$ 13/14) are reported in Figure \ref{evta}. The extension
of the pockets decreases with time, the first one beeing the largest.

\begin{figure}[h]
   \begin{center}
   \includegraphics[width=8cm]{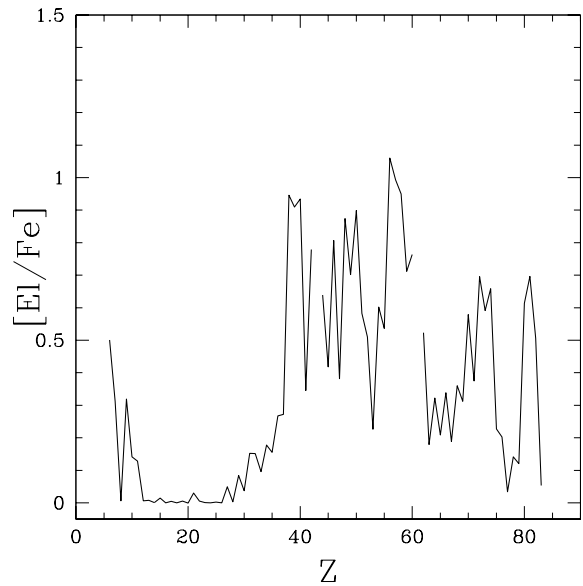}
   \end{center}
      \caption{Final surface composition of the model in Figure 1.
      The standard spectroscopic notation,
      namely [El/Fe]=log($N$(El)/$N$(Fe))$-$log($N$(El)/$N$(Fe))$_\odot$,
       is used, where El is the name of a generic element.}
         \label{surf}
\end{figure}

\begin{figure}[h]
   \begin{center}
   \includegraphics[width=9cm]{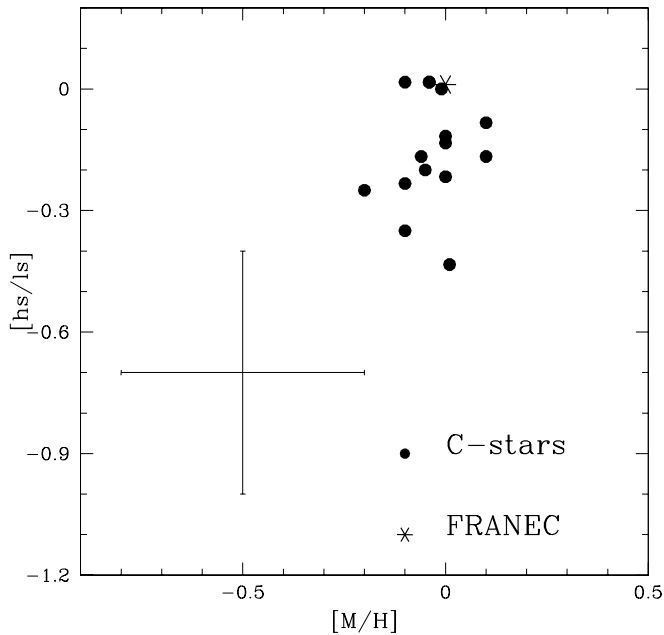}
   \end{center}
      \caption{Observed [hs/ls] ratio of a sample of galactic 
       C(N) stars, compared with our theoretical prediction. The huge cross represents 
       typical observational error bars of C(N) stars, whose spectra are very crowded (see \cite{ab02}).}
         \label{cstar}
\end{figure}

The ashes of the neutron-capture nucleosynthesis allowed by the $^{13}$C burning are spread
within the intershell by the subsequent convective thermal pulse. Later on, as a consequence of the 
third dredge up, the envelope  composition is  polluted with the
products of the s process. The final surface composition resulting after nine TDU episodes is 
reported in Figure \ref{surf}.
Among the light elements, note the logarithmic enhancements of C (0.5 dex), F (0.3 dex), Ne (0.14 dex) and Na (0.12 dex).
We recall that C is the main product of the $3\alpha$-burning. Fluorine is
synthesized in the convective shell generated by the thermal pulse
via the $^{15}$N$(\alpha,\gamma)^{19}$F reaction, the $^{15}$N being 
accumulated in the
$^{13}$C pocket mainly via the chain 
$^{14}$N$(n,p)^{14}$C$(\alpha,p)^{18}$O$(p,\alpha)^{15}$N\footnote{A minor
contribution to the synthesis of
the $^{15}$N in the pocket may eventually come
from the $^{14}$N$(n,\gamma)^{15}$N and from the $^{14}$N$(p,\gamma)^{15}$O reactions.}.
The nitrogen enhancement in the envelope (0.3 dex) is 
totally due to the first dredge up. In fact,
the $^{14}$N 
left by the H burning within the intershell is fully
converted into $^{22}$Ne during the convective thermal pulse,
giving rise to the resulting Ne enhancement. 
The Na enhancement
is partly due to the proton captures on $^{22}$Ne, occurring within the H-burning shell, and partly due to the neutron captures on $^{22}$Ne followed by the $^{23}$Ne decay, 
taking place within the
He intershell.  
Concerning the s-process nucleosynthesis, all the elements (from Sr to Pb) are enhanced at various degree.
We find that the abundance
of Sr, Y and Zr at the first s peak, the so-called ls elements ({\it light} s elements),
is comparable with the one of the hs elements ({\it heavy} s elements) Ba, La, Ce, Pr, Nd at the
second s peak. 
Lead is underproduced with respect to barium,
as expected for AGB stars of this metallicity (see next section).
In Figure \ref{cstar}, the predicted
[hs/ls] when (C/O=1) is compared to those measured in galactic C(N) giants \cite{ab02}.
Note that, for a given metallicity, this intrinsic index is indicative
of the physical conditions of the s-process site.

\section{Nucleosynthesis in low-metallicity AGB stars: lead stars}
\label{seclead}
The last challenging chapter of the AGB nucleosynthesis story is provided by the
high resolution spectroscopy of very metal-poor C- and s-rich stars. 
The present generation of halo stars is old ($\sim 14$ Gyr) 
and, therefore, is made of
low mass objects ($M$ $<$ 0.9 $M_\odot$). 
Then, when a low mass star reaches the AGB, its envelope is so 
small that the TDU never takes place. 
However, available spectroscopic surveys for very metal-poor stars 
(\cite{be99}, \cite{ch03}) found
that $\sim$20\% to 30\% of the candidates ([Fe/H] $>$ $-$2.5) are carbon rich.      
   \begin{figure}[h]
   \begin{center}
   \includegraphics[width=8cm,angle=-90]{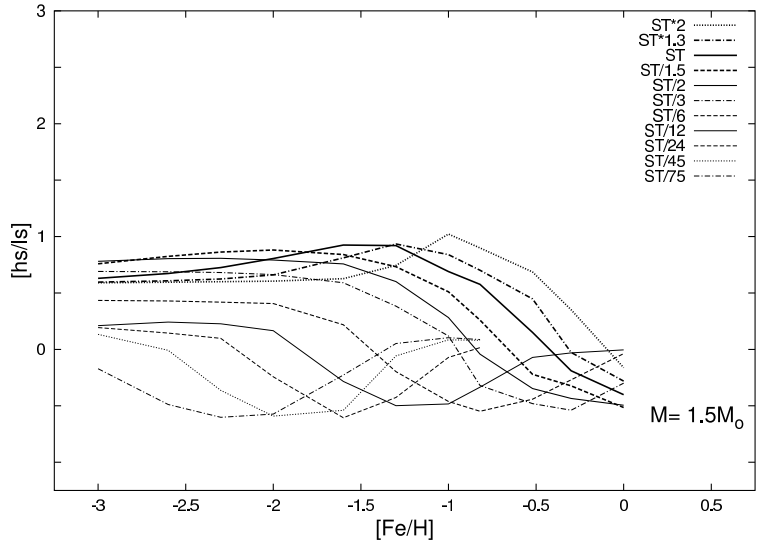}
   \end{center}
      \caption{Predicted [hs/ls] versus [Fe/H] for different 
      $^{13}$C pocket choices.}
         \label{del1}
   \end{figure}
These stars are probably low mass dwarfs or giants, 
with a lifetime comparable with the age of the Galaxy,
belonging to binary systems. It is possible that 
in most cases the C enhancement is the result of an 
ancient accretion process (by stellar wind or Roche lobe 
overflow \cite{jo92})
from a more massive AGB companion (now a white dwarf). 
In such a case, also the products 
of the neutron capture nucleosynthesis were enhanced in the material accreted onto 
the secondary star.

In low mass AGBs, the main source of neutrons, the \ctan,  
is primary-like, i.e. not directly affected by the 
metallicity of the pristine material. 
Nevertheless, the iron seeds scale with the metallicity, 
so that the lower the metallicity the larger is the number 
of neutrons available per seed. As a result, 
at low metallicities, most of the seeds are converted 
into $^{208}$Pb, at the termination point of the s-process 
fluence. Note that in stars suffering recurrent TDU episodes, the
amount of $^{12}$C in the envelope increases with time.
This primary carbon is mainly converted into $^{14}$N, 
within the H-burning shell, and, later on, into $^{22}$Ne, within 
the convective zone generated by a thermal pulse (see section \ref{secnsou}).
In very metal poor stars,
the primary neon accumulated in the intershell
becomes the main seed for the s process \cite{chief} (see also \cite{sg}).      
For these reasons, in a very metal-poor AGB star experiencing a few 
TDU episodes, a consistent enhancement of lead is expected \cite{bgw99}.
High-resolution spectroscopy of very metal-poor C-rich 
stars largely confirms such a qualitative expectation.
However, the precise prediction of the heavy element 
enhancements at the surface of the secondary star is more 
complex than previously sketched.
We have to recall that we are looking at the intershell 
material that was mixed with the envelope of the 
primary AGB stars during the various TDU episodes and, 
later on, further diluted
within the pristine material of the envelope of the 
secondary star. If this secondary star
(actually the C-rich object presently detected), is 
evolved off the main sequence, 
this late dilution may be particularly efficient. 
In addition, the amount of mass accreted
depends on the orbital parameters of the system 
that are, in most cases, unknown.

\begin{figure}[h]
   \begin{center}
   \includegraphics[width=8cm,angle=-90]{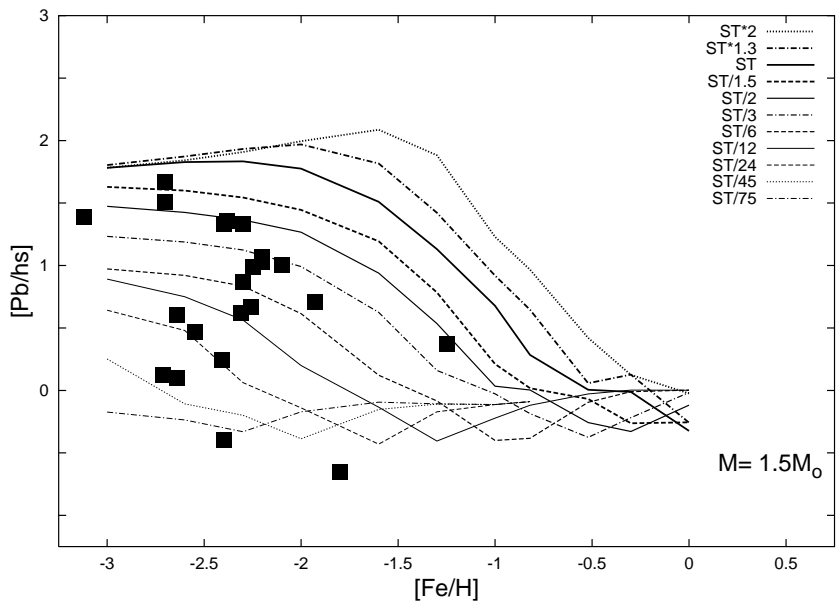}
   \end{center}
      \caption{Predicted [Pb/hs] versus [Fe/H] for different 
  $^{13}$C pocket choices. Spectroscopic data of s-rich 
  and lead-rich stars are included for comparison (adapted from \cite{del04}).}
         \label{del3}
   \end{figure}

In this context, in order to investigate the efficiency of the s-process site,
intrinsic indices [hs/ls] and [Pb/hs] are particularly useful
(\cite{ga04}, \cite{del04}). 
The variation of these two indices with metallicity are
shown in Figures \ref{del1} and \ref{del3}. 
These calculations are based on a grid  
of older FRANEC models, where the Reimers's parameterisation
($\eta$ = 0.5) was used and no velocity profile below the 
convective envelope was introduced. The n-capture nucleosynthesis 
was calculated with a post-process code that follows
in detail the physical and thermodynamical structure of 
the He-rich intershell in the TP-AGB phase. 
After each third dredge up, diffusion of protons in a thin 
layer in the top region of the He-rich intershell was assumed.
It gives rise to a $^{13}$C pocket at H reignition; the amount 
of $^{13}$C in the pocket and its profile in mass was 
taken as a free parameter.
The standard case (ST, see \cite{ga98}) was shown to 
reproduce the main s-process component in the solar system 
for low mass AGB stars and half-solar metallicity.
The $^{13}$C pocket efficiency was assumed 
identical for all thermal pulses followed by TDU. Neutrons released by the \ct~pocket are of 
primary origin, depending on proton captures on the freshly 
nucleosynthesised $^{12}$C.
Varying the metallicity, the s-process distribution  
changes because it depends on the number of neutron 
captured per seed and, at low metallicities, peaks
at Pb, at the termination of the s-process path.
As recalled in section \ref{neusc}, 
a large spread of $^{13}$C pocket efficiencies
is needed in order to explain the spectroscopic 
observations of MS-S-C(N) stars belonging to the galactic disk \cite{bu01}. In Figures 
\ref{del1}, \ref{del3}, different lines 
correspond to different
$^{13}$C pockets, where the ST case was multiplied or 
divided by a given factor as indicated in the inset.
The ST$\times$2 case is an upper limit because ingestion 
of a larger abundance of protons would favour the
$^{14}$N production by proton capture on $^{13}$C.
As a comparison, spectroscopic measurements of C and s-rich 
metal-poor stars are reported
in Figure \ref{del3}. It results that for very metal-poor stars, 
[hs/ls] is weakly affected by a change of 
the efficiency of the $^{13}$C pocket, while large spreads of [Pb/hs] are expected,
in quite satisfactory agreement with observations. This is in accord with the large 
spread of \ct~pocket efficiencies required by the observation of MS-S-C(N) stars (\cite{bu01}, \cite{ab01} and \cite{ab02}).

Very low metallicity s-rich stars have an additional peculiarity: they are usually 
also N-rich.
The enhancement of nitrogen could be due to the hot bottom burning,
 but this would exclude low mass AGB stars that have a too cool temperature at the base of the 
 convective envelope. The activation of 
the so-called cool bottom process during the AGB \cite{no03}, a slow deep circulation
taking place in the radiative region located between the H-burning 
shell and the inner border of the convective envelope,
could contribute to the systematic occurrence of the
nitrogen overabundance. 
More likely, \cite{ho90} (see also \cite{fu00}) 
suggested that the scarcity of CNO in  very metal-poor stars 
may allow the convective zone generated by the first strong TP
to ingest some protons from the envelope. In this case, a
hot CN cycle would take place within the He-rich intershell and consistent amounts of $^{13}$C and
$^{14}$N are produced. More recently \cite{iw04},
 on the basis of detailed stellar models computations,
proposed that a peculiar s-process nucleosynthesis, characterized by 
a high neutron density,
could be activated by $\alpha$-capture on the freshly synthesized $^{13}$C.
The detailed s-process nucleosynthesis for a low-mass and very metal-poor AGB model 
has been calculated by \cite{stra04}. We found that
 the ingestion of proton into the convective intershell region and the subsequent deep TDU 
 provide the required mechanism to explain the N enhancement, but 
the high neutron density is maintained only for a very short time (few days),
 with negligible consequences on the overall surface overabundances.

\section{Conclusion.}
\label{secconc}
We have reviewed the 
theoretical investigation of nucleosynthesis and evolution of low mass AGB stars. 
By introducing a full network (from H to Bi) in our stellar evolution code, we have
obtained a self-consistent nucleosynthesis calculation for a 2 $M_\odot$ stellar model with
solar composition.  

Below the convective envelope of a thermally pulsing AGB star,
the H burning, the He burning and the s process
modify the internal composition. Then, the occurrence of recurrent third dredge up
 episodes
induces the modification of the surface composition. 
The adoption of an exponential decay 
of the average convective velocity at the inner border of the convective
envelope allows us to follow the formation 
of a $^{13}$C pocket in the top layer of the He-rich intershell. 
During the interpulse, this pocket
is compressed and heated, untill neutrons are released via the \ctan~ reactions.   
This is the main neutron source, responsible for the s-process nucleosynthesis
 in low mass AGB stars.
A second (marginal) neutron burst, via the \nean~ reactions, may eventually occurs 
at the base of the convective zone generated by
a thermal pulse. One of the main source of uncertainty in modeling AGB stars concerns the choice of the
mass loss rate. 
A new semiempirical calibration of the mass loss has been derived from  a collection of 
the most recent observational data of variable AGB stars. This calibration has been used
for the AGB model calculation presented here. 
We have followed the evolution up to the last third dredge up episode.  
The final elemental distribution is representative of the one expected for the intrinsic 
carbon stars observed in the disk of the Milky Way. A comparison with available 
spectroscopic analysis shows a reasonable agreement.

40 years have passed since the discovery of the thermal pulses by
Schwarzschild \& H\"arm \cite{scha65} and 
many efforts have been made for the comprehension of the AGB evolution and nucleosynthesis.
The progressive improvement of the computational capability allows
the development of more complex theoretical tools. The main tasks for the next future will be
the comprehension of the role played by hydrodynamical phenomena usually neglected in standard 
stellar model computations (see e.g
\cite{dsti96}, \cite{la99}, \cite{no03}, \cite{bu04}, \cite{hw03}, \cite{hw04}) and the
refinement of the nuclear physics input data.           

We wish to express our thanks to Debora Delaude and Marco Pignatari for helping in the
preparation of this manuscript. We are indebted to Franz K\"appeler and 
Inma Dom\'\i nguez for
a long profitable and continuous interaction.
Part of this work has been supported by the Italian MIUR-FIRB Project "The Astrophysical 
Origin of the Heavy Elements beyond Iron" and by the Italian MIUR-PRIN Project 2004 "Nuclear astrophysics in low mass stars".          

\bibliographystyle{plain}

\end{document}